\documentclass[useAMS,fleqn,usenatbib]{mnras}
\pdfoutput=1 %for arxiv
\usepackage{graphicx}
\usepackage{color}
\usepackage{amsmath,amssymb,amstext}
\usepackage[T1]{fontenc}
\usepackage{newtxtext,newtxmath}
\usepackage[figure,figure*]{hypcap}
\usepackage[most]{tcolorbox}
\input{hyperlink-year-only-natbib-patch}

\tcbset{
    frame code={}
    center title,
    left=0pt,
    right=0pt,
    top=0pt,
    bottom=0pt,
    colback=gray!70,
    colframe=white,
    width=42pt,
    enlarge left by=0mm,
    boxsep=0pt,
    arc=0pt,outer arc=0pt,
    }

\usepackage[normalem]{ulem}

\input{macros_dc.sty}

\definecolor{BurntOrange}{RGB}{207, 83, 0}

\title[The Galaxy Clustering Crisis]{The Galaxy Clustering Crisis in Abundance Matching}
\author[Campbell et al.]{%
Duncan Campbell$^{1}$\thanks{E-mail:  \mailto{duncan.campbell@yale.edu}},
Frank C. van den Bosch$^{1}$,
Nikhil Padmanabhan$^{2, 3}$,\newauthor 
Yao-Yuan Mao$^{4}$, 
Andrew R. Zentner$^{4}$,
Johannes U. Lange$^{1, 5}$,
Fangzhou Jiang$^{6}$,\newauthor 
and Antonio Villarreal$^{4}$\vspace*{0.2em} \\ 
$^{1}$Department of Astronomy, Yale University, New Haven, CT 06511, USA\\
$^{2}$Department of Physics, Yale University, New Haven, CT 06520, USA \\
$^{3}$Yale Center for Astronomy and Astrophysics, Yale University, New Haven, CT 06520, USA\\
$^{4}$Department of Physics and Astronomy \& Pittsburgh Particle Physics, Astrophysics, and Cosmology Center (PITT PACC),\\ University of Pittsburgh, Pittsburgh, PA 15260, USA\\
$^{5}$Kavli Institute for Theoretical Physics, University of California, Santa Barbara, CA 93106, USA\\
$^{6}$Racah Institute of Physics, The Hebrew University, Jerusalem 91904, Israel
}

\pubyear{2017}

\begin{document}
\label{firstpage}
\pagerange{\pageref{firstpage}--\pageref{lastpage}}
\maketitle

\begin{abstract}
Galaxy clustering on small scales is significantly under-predicted by sub-halo abundance matching (SHAM) models that populate (sub-)haloes with galaxies based on peak halo mass, $M_{\rm peak}$.  SHAM models based on the peak maximum circular velocity, $V_{\rm peak}$, have had much better success.  The primary reason $M_{\rm peak}$ based models fail is the relatively low abundance of satellite galaxies produced in these models compared to those based on $V_{\rm peak}$.  Despite success in predicting clustering, a simple $V_{\rm peak}$ based SHAM model results in predictions for galaxy growth that are at odds with observations.  We evaluate three possible remedies that could ``save'' mass-based SHAM: (1) SHAM models require a significant population of ``orphan'' galaxies as a result of artificial disruption/merging of sub-haloes in modern high resolution dark matter simulations; (2) satellites must grow significantly after their accretion; and (3) stellar mass is significantly affected by halo assembly history. No solution is entirely satisfactory. However, regardless of the particulars, we show that popular SHAM models based on $M_{\rm peak}$ cannot be complete physical models as presented.  Either $V_{\rm peak}$ truly is a better predictor of stellar mass at $z\sim 0$ and it remains to be seen how the correlation between stellar mass and $V_{\rm peak}$ comes about, or SHAM models are missing vital component(s) that significantly affect galaxy clustering.
\end{abstract}

\begin{keywords}
galaxies: halos -- galaxies: evolution -- galaxies: clustering
\end{keywords}

\section{introduction}

The prevailing picture of galaxy formation is intricately tied to that of dark matter structure formation.  The matter distribution of the Universe evolved from a relatively smooth state into a complex web of structure over $\sim 14$ billion years. Within this web, small inhomogeneities evolved into extended, quasi-spherical, gravitationally bound objects called haloes. The build-up of structure proceeds hierarchically as smaller haloes are accreted into larger haloes, becoming substructures called sub-haloes. The potential wells of dark matter (sub-)haloes are the natural sites for galaxies to form as baryons cool and condense into stars \citep{Rees:1977gr, White:1978uk, Fall:1980br}. 

This general, albeit coarse, view of galaxy formation fits well with the premise of sub-halo abundance matching (SHAM).  SHAM in its most simple form rests on the hypothesis that all massive\footnote{above a limiting lower mass scale below which galaxy formation becomes inefficient} (sub-)haloes host galaxies, all galaxies occupy (sub-)haloes, and there is a simple monotonic relation between galaxy mass and the mass of the (sub-)halo each galaxy occupies \citep{Kravtsov:2004fi, Vale:2004bb, Conroy:2006iz}.  This approach relies on simulations of dark matter structure formation to provide robust statistical predictions for many properties of dark matter (sub-)haloes, e.g. halo mass functions, mass profiles, and the spatial clustering of haloes.  By ``populating'' dark matter simulations with galaxies using the SHAM technique, the statistical predictions from simulations can be leveraged to constrain the galaxy-halo connection and make predictions for how galaxies form and evolve along with (sub-)haloes \citep{Springel:2006fq}. 

The stellar mass-halo mass (SMHM) relation is one of the most fundamental implications of SHAM.  The SMHM relation at redshift $\sim 0$ inferred using SHAM is consistent with HOD/CLF analysis \citep{Yang:2003du, vandenBosch:2003fd, vandenBosch:2007ea, Yang:2013hw, Zu:2015tt} and with independent, more direct, measurements from galaxy--galaxy lensing, satellite kinematics, and the Tully--Fisher relation \citep{vandenBosch:2004mn, Wang:2006fe, More:2009ws, Guo:2010do, Wang:2010ho, Moster:2010ep, Behroozi:2010ja, Mandelbaum:2015wt, Desmond:2015gr, vanUitert:2016fy}.  By extending this analysis to higher redshifts, the inferred evolution of the SMHM relation and the mass growth histories of haloes predicted by simulations provide constraints on the average stellar mass growth histories of galaxies that are consistent with the cosmic star-formation history of the Universe \citep{Conroy:2009ba, Wang:2010ho, Yang:2012ew, Moster:2013ab, Behroozi:2013fg, RodriguezPuebla:2017uo}.  An important finding of these studies is that the peak in star-formation efficiency shifts towards more massive haloes at earlier times. 

\begin{figure}
    \includegraphics[width=\columnwidth]{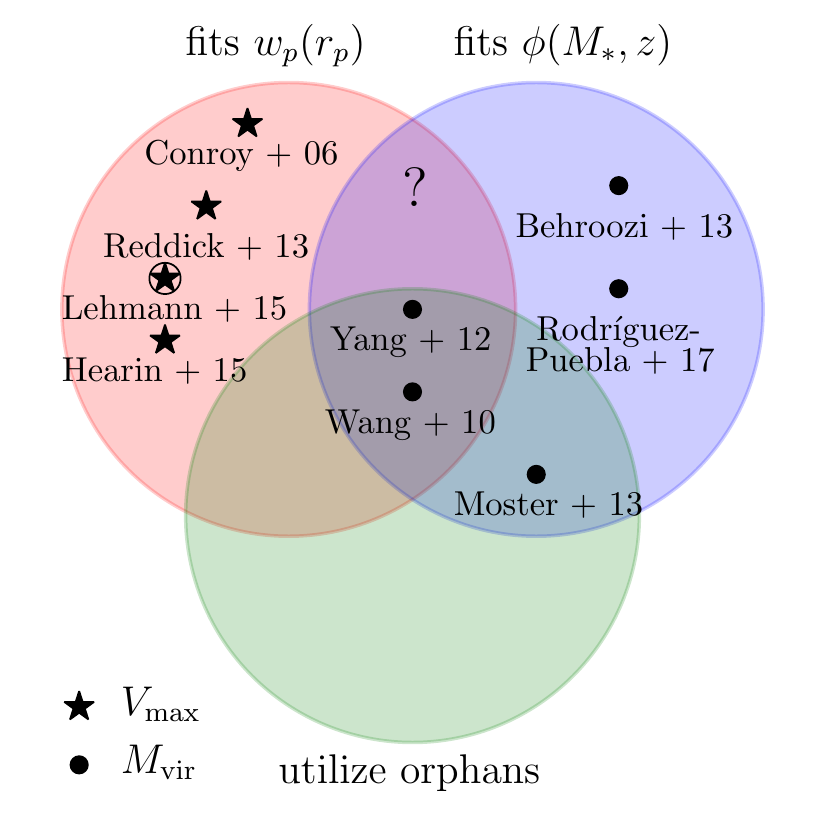}
    \caption{Venn diagram of selected SHAM models within three sets; those that fit the galaxy clustering, $w_p(r_p)$ (upper left red region), those that fit the evolution of the galaxy stellar mass function, $\phi(M_*,z)$ (upper right blue region), and those that utilize orphan galaxies (lower green region).  Models marked with a circle are based on measures of halo mass, $M_{\rm vir}$, including peak quantities.  Models marked with a star are based on measures of halo maximum circular velocity, $V_{\rm max}$, including peak quantities, with the \citet{Lehmann:2017fy} model falling somewhere in between.   This paper focuses on the lack of models in the region marked with a ``?''. }
    \label{fig:venn_diagram}
\end{figure}

The ability of SHAM to accurately predict the clustering of galaxies is more limited.  In particular, the small scale clustering of galaxies remains difficult to fit.  SHAM has only been shown to be consistent with galaxy clustering under one of two conditions: either stellar mass is tied to the peak circular velocity of (sub-)haloes (or closely related quantities) \citep{Reddick:2013gi, Hearin:2014hh, Lehmann:2017fy}, or the abundance of sub-haloes is treated as a free parameter \citep{Wang:2006fe, Wang:2010ho, Guo:2010do, Moster:2010ep, Yang:2012ew}.  This second approach relies on a population of ``orphan'' galaxies which have no identifiable sub-halo in a simulation.  As we will show in this work, both of these assumptions are problematic.  The assumption that stellar mass should be better correlated with maximum circular velocity over halo mass has not been robustly motivated.  Furthermore, we show that models that make this assumption are at odds with previous work on the evolution of the SMHM relation.  Orphan galaxies are difficult to reconcile with modern high resolution dark matter simulations which aim to resolve substructure.  We summarize this state of affairs in Fig. \ref{fig:venn_diagram}.  There is no published SHAM model which fits both galaxy clustering and the evolution of the stellar mass function using resolved substructure in a dark matter simulation without the addition of ``orphan'' galaxies.

The goal of this paper is threefold: (i) to make explicit the tension between fitting galaxy clustering measurements and the evolution of the stellar mass function within the SHAM framework; (ii) to examine the successes and failures of SHAM models based on maximum circular velocity; and (iii) to evaluate mechanisms to alleviate the clustering crisis in models based on halo mass.  The structure of this paper is as follows: In \S \ref{sec:methods} we describe the SHAM models used in this work and our implementation of each.  In \S \ref{sec:clustering} we examine the galaxy clustering signal prediction of each model.  In \S \ref{sec:vpeak_miracle} we describe how it is that SHAM based on $V_{\rm peak}$ is able to fit clustering while models based on $M_{\rm peak}$ generally do not.  In \S \ref{sec:not_such_a_miracle} we discuss some pitfalls of $V_{\rm peak}$ SHAM and explicitly define the "crisis" referenced in the title of this paper.  \S \ref{sec:orphans}--\S \ref{sec:assem_bias} go through physically motivated alterations to $M_{\rm peak}$ based SHAM models which alleviate the clustering crisis to various degrees.  We provide a summary and discussion of the main conclusions of this paper in \S \ref{sec:discussion}.

The models and analysis in this paper utilize the framework and code base in {\tt Halotools} \citep{Hearin:2016tc}, an {\tt Astropy} \citep{TheAstropyCollaboration:2013cd} affiliated Python\footnote{\http{www.python.org}} package.  We also make available all of the code and data products necessary to reproduce the figures and analysis in this paper\footnote{\https{github.com/duncandc/Clustering\_Crisis}}.  Throughout, we scale all units by $h = H_0/[100~{\rm km/s}~{\rm Mpc}^{-1}]$ where appropriate, and we use $\log(x)$ to indicate the base 10 logarithm of $x$.

\section{Methods}
\label{sec:methods}

\subsection{Abundance Matching Models}

SHAM, in its simplest form, assumes that the cumulative abundance of galaxies and haloes can be used to map galaxy properties uniquely onto (sub-)haloes.  Using stellar mass, $M_{*}$, and a halo mass proxy $\mathcal{M}$, SHAM assumes:
\begin{equation}
\label{eq:sham}
N(>M_*) = N(>\mathcal{M}) \implies M_{*} = f(\mathcal{M})
\end{equation} 
where $f(\mathcal{M})$ is some monotonically increasing function.  This maps the most massive galaxies to the most massive haloes.  The function $f(\mathcal{M})$ may be determined non-parametrically by matching the rank orders of galaxies and haloes.  We refer to this method of SHAM as ``rank order'' abundance matching.  However, we also consider another class of SHAM models where $f(\mathcal{M})$ is parametrized.  This introduces the possibility that the observed abundance of galaxies is not strictly matched; however, it is generally the goal of such models to preserve this characteristic. Here we use the umbrella term, SHAM, to refer to both the rank order abundance matching and parametrized SMHM relation methods.

The predictions of SHAM are sensitive to the details of how it is implemented.  Many halo properties have been examined in search of which ``best'' correlates with galaxy stellar mass or luminosity.  It is generally found that ``peak'' values of mass-like properties, estimated over the history of the (sub-)halo, work best.  Specifically, using the peak maximum circular velocity reproduces the clustering statistics of galaxies most successfully \citep{Conroy:2006iz, Reddick:2013gi}.  The difference between the peak values and current values is most pronounced for sub-haloes, as sub-haloes are subject to stripping processes which remove mass, while the core of the sub-halo, which hosts a galaxy, is thought to survive much longer \citep[e.g.][and references therein]{Behroozi:2014bg, vandenBosch:2016cj, Jiang:2016dw, vandenBosch:2016ky}.  Implicit to the SHAM algorithm is that the processes that set the properties of central and satellite galaxies are independent of halo assembly history.  This assumption has been relaxed somewhat by allowing satellites to either grow or lose stellar mass after being accreted into a host-halo \citep{Watson:2012hw, Yang:2012ew,  Behroozi:2015wx} and is well motivated by observations \citep{Wetzel:2013dw}.  Furthermore, the very distinction between host-haloes and sub-haloes is blurred by the recognition that some haloes are accreted into a host-halo, only for their orbit to take them outside their host's virial radius.  For these ``backsplash'' haloes, it may be more appropriate to treat them as sub-haloes likely to host galaxies that have properties more in common with traditionally defined satellite galaxies \citep{Mamon:2004jy, Wetzel:2014up}.

Below, we review a set of SHAM models and our method of implementing each before examining the clustering predictions of each model in the following section.  We provide a summary of the models used in this paper in table 1.

\subsubsection{Rank Order SHAM}
\label{sec:rank_sham}

The simplest implementation of SHAM maps galaxies into (sub-)haloes by matching rank orders.  Given a set of galaxies, $M_*=\{M_{*i}\}$, and (sub-)haloes, $\mathcal{M}=\{\mathcal{M}_j\}$, of the same size and drawn from equivalent representative volumes, this method proceeds by calculating the ranks of each:
\begin{align}
n^{\rm gal} &= \mathcal{R}(M_*) \nonumber \\
n^{\rm halo} &= \mathcal{R}(\mathcal{M})
\end{align}
where the $\mathcal{R}(x)$ function returns the ordinal ranks.  Each galaxy is then assigned to a (sub-)halo with the equivalent rank, i.e. $n_i^{\rm gal}=n_j^{\rm halo}$.  We consider two rank order SHAM models in this paper, one based on $M_{\rm peak}$ and one on $V_{\rm peak}$.  As shorthand, we refer to these models as ``RM'' and ``RV'', respectively. 

Given a set of galaxies and haloes, this mapping is deterministic.  However, the assumption of perfect rank ordering with respect to the cumulative distribution can be relaxed, allowing for stochasticity in the mapping and resulting in a probabilistic relation between stellar mass and halo mass, $P(M_*|\mathcal{M})$.  There are various methods used in the literature to add scatter to this relation while maintaining agreement with an observed stellar mass function.  One may attempt to deconvolve the stellar mass function from the scatter model such that, after solving for eq. (\ref{eq:sham}) with this deconvolved function and applying the scatter model, the new stellar mass function is consistent with the observed function \citep[e.g.][]{Behroozi:2010ja}.  Another option is to manually add scatter to the stellar masses used in abundance matching, re-ranking on the new values to perform the matching, and iteratively solving for a solution that results in the desired amount of scatter in the SMHM relation \citep[e.g.][]{Hearin:2013ok}. 

We do not go through the additional step of adding scatter to our RM and RV models, as scatter in the SMHM relation generally \emph{decreases} the strength of the clustering signal for massive galaxies ($M_* > 10^{11} ~h^{-2} M_{\odot}$) with little effect at lower masses \citep{Tinker:2016vu}, and our work focuses on the \emph{decrement} in galaxy clustering signal for samples of less massive galaxies in SHAM models.

\begin{table}
\label{table:models}
\caption{Summary SHAM models used in this paper}
\begin{tabular}{ |p{0.55cm}|p{3cm}|p{3.65cm}}
\multicolumn{3}{|c|}{} \\
\hline
name & reference & description  \\
\hline
\hline
RM &  -- & rank order SHAM on $M_{\rm peak}$ \\
RV & -- & rank order SHAM on $V_{\rm peak}$ \\
M13 & \citet{Moster:2013ab} & evolving SMHM model \\
B13 & \citet{Behroozi:2013fg} & evolving SMHM model \\
Y12 & \citet{Yang:2012ew} & evolving SMHM model \\
\hline
\end{tabular}
\end{table}

\subsubsection{Moster et al. (2013) SHAM}
\label{sec:m13}

A distinct method from the rank order SHAM method is to parametrize the SMHM relation, fitting for the parameters which result in a stellar mass function that is consistent with an observed function after populating a dark matter simulation.  The first model of this type we consider is by \cite{Moster:2013ab}. In this model, the SMHM relation is parametrized as a function of halo mass and redshift to account for evolution.  Furthermore, it is assumed that the stellar mass of satellite galaxies is determined by the mass of sub-haloes at the time of accretion into a more massive host-halo for the first time.  The functional form is given by:
\begin{equation}
\label{eq:moster_smhm}
\frac{\langle M_{*} | \mathcal{M}\rangle_{\rm med}(a)}{\mathcal{M}} = 2N\left[ \left(\frac{\mathcal{M}}{{M}_1} \right)^{-\beta} + \left( \frac{\mathcal{M}}{{M}_1} \right)^{\gamma} \right]^{-1}
\end{equation}
where $\langle\rangle_{\rm med}$ indicates the median.  The evolution of the parameters is given by:
\begin{align}
\label{eq:smhm_params}
\log[{M}_1(a)]& = {M}_{10} + {M}_{11}(1-a) \\
N(a) &= N_{10} + N_{11}(1-a) \\
\beta(a) &= \beta_{10} + \beta_{11}(1-a) \\ 
\gamma(a) &=  \gamma_{10} + \gamma_{11}(1-a)
\end{align}
where the scale factor is either the instantaneous scale factor, or the one at the time of accretion:
\begin{equation}
a = 
\begin{cases}
    a      & \quad \text{if host-halo}\\
    a_{\rm acc}  & \quad \text{if sub-halo} \\
  \end{cases}
\end{equation}
and $\mathcal{M}$ is the instantaneous mass or the mass at accretion, $M_{\rm acc}$, for sub-haloes.  Scatter in the SMHM relation at fixed halo mass is modelled as a fixed log-normal with $\sigma_{\log(M*)} \approx 0.18$.

\citet{Moster:2013ab} find the best fit parameters for equation (\ref{eq:moster_smhm}) that reproduce the stellar mass function at various redshifts between $z=0$ and $z \sim 4$.  They also show that the implied star-formation rates of galaxies given the growth history of the (sub-)haloes in their simulation is consistent with the cosmic star-formation rate density.  The parameters were constrained using the Millennium simulation with (sub-)haloes defined as spherical over-densities (SO) with mean internal density 200 times the critical density of the universe, $M_{200c}$.   As shorthand, we will refer to this model as ``M13''.  The parameter values used in this paper are taken directly from \citet{Moster:2013ab} and listed in table 2.
      
\subsubsection{Behroozi et al. (2013) SHAM}
\label{sec:b13}

\citet{Behroozi:2013fg} make similar assumptions as M13 but utilize a different parametrization given by:
\begin{equation}
\log[\langle M_*|\mathcal{M} \rangle_{\rm med}(a)] = \log(\epsilon M_1) + f(\log(\mathcal{M}/M_1)) - f(0)
\end{equation}
where,
\begin{equation}
f(x) = -\log(10^{\alpha x}+1) + \delta\frac{\left[ \log(1+\exp[x])\right]^{\gamma}}{1+\exp(10^{-x})}
\end{equation}
and where $\mathcal{M} = M_{\rm peak}$.  The parameters evolve with redshift as:
\begin{align}
\nu(a) & = e^{-4a^2}  \\
\log[M_1(a)] & =  M_{10} + \nu\left[ M_{1,a}(a-1) + M_{1,z}z \right] \\
\log[\epsilon(a)] & = \epsilon_0 + \nu \left[\epsilon_a(a-1) +\epsilon_z z\right] + \\ \nonumber 
& ~~~~ \epsilon_{a,2}(a-1)  \\
\alpha(a) & =  \alpha_{0} + \nu\left[ \alpha_a(a-1)\right] \\
\delta(a) & =   \delta_{0} + \nu\left[ \delta_a(a-1) + \delta_z z\right] \\
\gamma(a) &= \gamma_{0} + \nu\left[ \gamma_a(a-1) + \gamma_z z\right]
\end{align}
When fitting their model, \cite{Behroozi:2013fg} also allow the scatter in the SMHM relation to vary as a function of redshift.  The variation found is consistent with no variation (constant over cosmic time), so we neglect the small variation in the best fit model and use a constant non-varying scatter.  We have checked that including this variation has no appreciable effect on our conclusions. \cite{Behroozi:2013fg} constrained the parameters of this model by fitting the stellar mass function, specific star-formation rates, and the cosmic star-formation history between z=0-8.  The values used in this paper are taken directly from \cite{Behroozi:2013fg} and listed in table 2. As shorthand, we will refer to this model as ``B13''.

\subsubsection{Yang et al. (2012) SHAM}
\label{sec:y12}

\cite{Yang:2012ew} take a different approach than M13 and B13.  They allow for evolution with redshift in a similar manner, but additionally allow satellites to grow or lose mass after the time of accretion.  The SMHM relation for central galaxies is: 
\begin{equation}
\langle M_{*,\rm cen} |  \mathcal{M} \rangle_{\rm med}(z) = M_{0}\frac{\left( \mathcal{M}/M_1 \right)^{\alpha+\beta}}{\left(1+ \mathcal{M}/M_1 \right)^{\beta}}
\end{equation}  
where $\mathcal{M}=M_{180b}$, and the parameters\footnote{We have altered the naming scheme somewhat to be more consistent with the other models.} evolve with redshift as:
\begin{align}
\log[M_{0}(z)] &= M_{01} + \gamma_1 z \\
\log[M_{1}(z)] &= M_{11} + \gamma_2 z \\
\alpha(z) &= \alpha_0 + \gamma_3 z \\
\log[\beta(z)] &= \log(\beta_0) + \gamma_4 z + \gamma_5 z^2
\end{align}
The median stellar mass of satellites is determined by interpolating between the stellar mass a satellite has at the time of accretion and the stellar mass of a central galaxy with the same halo mass as the satellite at accretion at redshift, $z$.  This value is determined by a single parameter, $c$:
\begin{align}
\label{eq:y12sat_model}
\langle M_{*,\rm sat}(z, z_{\rm acc})\rangle =&  (1-c)\langle M_{*,\rm cen} |  \mathcal{M}_{\rm acc} \rangle(z_{\rm acc})~+ \\ \nonumber &  c\langle M_{*,\rm cen}|  \mathcal{M} \rangle(z)
\end{align}
where $\mathcal{M}_{\rm acc}$ is the halo mass proxy at the time of accretion for the sub-halo.  The case where $c=0$ corresponds to no growth (or mass loss) since $z_{\rm acc}$.  The case where $c=1$ corresponds to using the same SMHM relation for centrals and satellites at all redshifts (i.e., akin to standard SHAM).  We discuss post-accretion satellite evolution, including this model, in more detail in \S \ref{sec:sat_growth}.

Apart from the parametrization of the SMHM relation and its evolution, the original implementation of \cite{Yang:2012ew} differs significantly from the previous two models in that it uses a fully analytical halo and sub-halo model for abundances, sub-halo profiles, and halo bias \citep[see][]{Yang:2011ou}.  We take a different approach and use the SMHM relation of the model and apply it directly to a simulation, side-stepping the need to analytically model these components.  Multiple fits were performed in \citet[][]{Yang:2012ew}.  We use the parameter constraints determined by fitting the stellar mass function at multiple epochs (z=0-5) as well as the conditional stellar mass function at $z \sim 0$.  This set of parameters was also shown to fit galaxy clustering observations well at $z\sim 0$.  The values used in this paper are listed in table 2, specifically these are taken from Table 4 (ID=4) in \citet[][]{Yang:2012ew}.  As shorthand, we will refer to this model as ``Y12''.

\begin{table}
\label{table:model_parameters}
\caption{Parameter values used for the evolving SHAM models presented in this paper. Mass parameters in M13 and B13 are scaled to $h=0.7$, while in Y12, $h=1$ as is the practice in the rest of this paper.  We use the parameters as is, and scale the output stellar masses to $h=1$.}
\begin{tabular}{ |p{1.5cm}|p{1.6cm}|p{1.2cm}|p{2cm}}
\multicolumn{4}{|c|}{} \\
\hline
model & parameter & value & uncertainty \\
\hline
\hline
M13		& $M_{10}$ 		&  11.590 		& $\pm~0.236$ 			\\%& $\log(M_{\odot})$ \\
M13 		& $M_{11}$ 		&  1.195  		& $\pm~0.353$			\\%& $\log(M_{\odot})$ \\	
M13 		& $N_{10}$ 		&  0.0351 		& $\pm~0.0058$			\\%& -- \\
M13 		& $N_{11}$ 		&  -0.0247		& $\pm~0.0069$			\\%& --  \\
M13 		& $\beta_{10}$ 		&  1.376 		& $\pm~0.153$			\\%& --  \\
M13 		& $\beta_{11}$ 		& -0.826  		& $\pm~0.225$ 			\\%& -- \\
M13 		& $\gamma_{10}$ 	&  0.608 		& $\pm~0.059$			\\%& --  \\
M13 		& $\gamma_{11}$ 	&  0.329 		& $\pm~0.173$ 			\\%& -- \\
\\
B13		& $M_{10}$ 		&  11.514 		& $\pm^{0.053}_{0.009}$ 	\\%& $\log(M_{\odot})$ \\
B13		& $M_{1,a}$ 		&  -1.793 		& $\pm^{0.315}_{0.330}$ 	\\%& $\log(M_{\odot})$ \\
B13		& $M_{1,z}$ 		&  -0.251		& $\pm^{0.012}_{0.125}$ 	\\%& $\log(M_{\odot})$ \\
B13		& $\epsilon_{0}$ 	&  -1.777		& $\pm^{0.133}_{0.146}$ 	\\%& -- \\
B13		& $\epsilon_{a}$ 	&  -0.006		& $\pm^{0.113 }_{0.361}$ 	\\%& -- \\
B13		& $\epsilon_{z}$ 	&  -0.000		& $\pm^{0.003}_{0.104}$ 	\\%& -- \\
B13		& $\epsilon_{a,2}$ 	&  -0.119		& $\pm^{0.061}_{0.012}$ 	\\%& -- \\
B13		& $\alpha_{0}$ 		&  -0.119		& $\pm^{0.061}_{0.012}$ 	\\%& -- \\
B13		& $\alpha_{a}$ 		&  -0.119		& $\pm^{0.061}_{0.012}$ 	\\%& -- \\
B13		& $\delta_{0}$ 		&  -1.777		& $\pm^{0.133}_{0.146}$ 	\\%& -- \\
B13		& $\delta_{a}$ 		&  -0.006		& $\pm^{0.113 }_{0.361}$ 	\\%& -- \\
B13		& $\delta_{z}$ 		&  -0.000		& $\pm^{0.003}_{0.104}$ 	\\%& -- \\
B13		& $\gamma_{0}$ 	&  -1.777		& $\pm^{0.133}_{0.146}$ 	\\%& -- \\
B13		& $\gamma_{a}$ 	&  -0.006		& $\pm^{0.113 }_{0.361}$ 	\\%& -- \\
B13		& $\gamma_{z}$ 	&  -0.000		& $\pm^{0.003}_{0.104}$ 	\\%& -- \\
\\
Y12		& $M_{01}$ 		&  10.36		& $\pm^{0.05}_{0.06}$	\\%& $\log(h^{-2}M_{\odot})$ \\
Y12		& $M_{11}$ 		&  11.06		& $\pm^{0.08}_{0.15}$	\\%& $\log(h^{-1}M_{\odot})$ \\
Y12		& $\alpha_{0}$ 		&  0.27 		& $\pm^{0.01}_{0.01}$	\\%& -- \\
Y12		& $\beta_{0}$ 		&  4.34		& $\pm^{0.96}_{0.52}$	\\%& -- \\
Y12		& $\gamma_{1}$ 	&  -0.96		& $\pm^{0.13}_{0.19}$ 	\\%& -- \\
Y12		& $\gamma_{2}$ 	&  -0.23		& $\pm^{0.05 }_{0.06}$ 	\\%& -- \\
Y12		& $\gamma_{3}$ 	&  -0.41		& $\pm^{0.07}_{0.08}$ 	\\%& -- \\
Y12		& $\gamma_{4}$ 	&  -0.11		& $\pm^{0.11}_{0.08}$ 	\\%& -- \\
Y12		& $\gamma_{5}$ 	&  0.01		& $\pm^{0.05 }_{0.07}$ 	\\%& -- \\
Y12		& $c$ 			&  1.0		& -- 					\\%& -- \\

\hline
\end{tabular}
\end{table}

\subsection{Dark Matter Simulations}
\label{sec:dm_sim}

We build mock galaxy catalogues using the 5 SHAM models described in the preceding sections using the Bolshoi \citep{Klypin:2011bd} simulation output at $z=0$.  The Bolshoi simulation follows the evolution of $2048^3$ dark matter particles using the Adaptive Refinement Tree (ART) code \citep*{Kravtsov:1997iy} in a flat $\Lambda$CDM cosmology with parameters $\Omega_{\rmm,0} = 1 - \Omega_{\Lambda,0} = 0.27$, ${\Omega}_{\rmb,0} = 0.0469$, $n_\rms = 0.95$, $\sigma_8 = 0.82$, and $h = 0.7$ (hereafter ``Bolshoi cosmology''). The box size of the Bolshoi simulation is $L_{\rm box} = 250 \mpch$, with a dark matter particle mass of $m_{\rm p} = 1.35 \times 10^8 \msunh$.

\begin{figure}
    \includegraphics[width=\columnwidth]{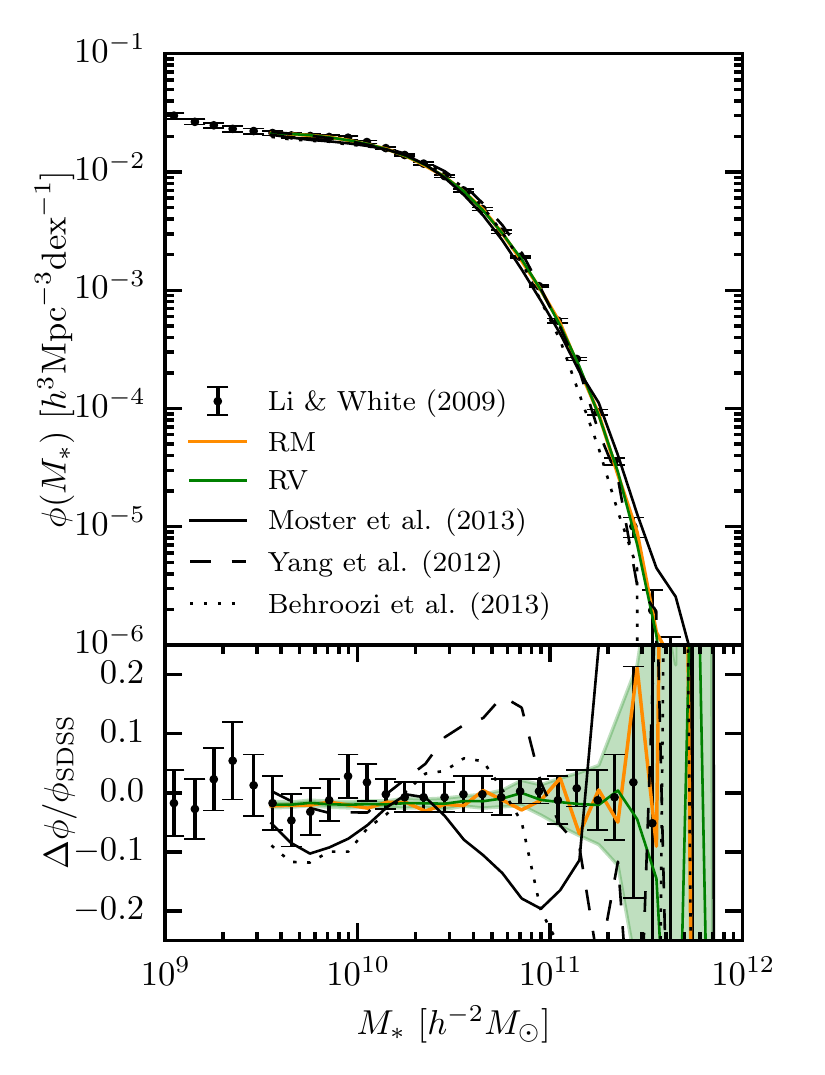}
    \caption{The stellar mass function of a mock realization of each model is plotted (lines).  For comparison we show the stellar mass function in SDSS as measured by \citet{Li:2009kh} as points with error bars.  The lower panel shows the residuals with respect to the triple Schechter fit from \citet{Li:2009kh}. As an example, we show the $\pm 1 \sigma$ error on the prediction for the RV model as the shaded green region, calculated by re-populating the simulation with the model $50$ times. While not shown, the uncertainty on the other model predictions are similar.}
    \label{fig:phi}
\end{figure}

(Sub-)haloes are found using the phase-space halo finder {\tt ROCKSTAR} \citep{Behroozi:2013cn}, which uses adaptive, hierarchical refinement of friends-of-friends groups in six phase-space dimensions and one time dimension, and tracked over time using the {\tt Consistent Trees} algorithm \citep{Behroozi:2013dz}.  As demonstrated in \cite{Knebe:2011jc, Knebe:2013bp}, this results in a very robust tracking of (sub-)haloes \citep[also see][]{vandenBosch:2016ky}. Haloes in this catalogue are defined to be spherical volumes centred on a local density peak (SO hereafter), such that the average density inside the sphere is $\bar{\rho}_{\rm h}(z) = \Delta_{\rm vir}(z)\rho_{\rm m}(z)$.  Here $\rho_{\rm m}(z) = \Omega_{\rm m}(z) \rho_{\rm crit}(z)$, where $\rho_{\rm crit}(z) = 3H(z)^2/8\pi G$ is the critical energy density of the Universe, and $\Delta_{\rm vir}(z)$ is given by a fitting function \citep{Bryan:1998cc}:
\begin{equation}
\Delta_{\rm vir}(z) = \left[ 18\pi^2 - 82\Omega_{\Lambda}(z) - 39\Omega_{\Lambda}^2(z) \right]\Omega_m^{-1}
\end{equation}
For the Bolshoi cosmology, $\Delta_{\rm vir}(z=0)\simeq 360$.  The radius of each such sphere defines the virial radius $R_{\rm vir}$ of the halo, which is related to the mass of the halo via $M_{\rm vir} = (4/3)\pi R_{\rm vir}^3\bar{\rho}_{\rm h}$. Additionally, sub-haloes in this catalogue are distinct, self-bound structures whose centres are found within the virial radius of a more massive host-halo.  For each (sub-)halo, the maximum circular velocity is defined as: $V_{\rm max} \equiv {\rm Max}[GM(<r)/r]$, where $M(<r)$ is the mass enclosed within a distance r of the (sub-)halo centre.

From this catalogue we construct our mocks primarily using three values for each (sub-)halo: $M_{\rm peak}$, $V_{\rm peak}$, and $z_{\rm acc}$.  $M_{\rm peak}$ is defined as the peak virial mass a (sub-)halo achieves over its history.  In our halo catalogues, we retain all (sub-)haloes that obtained a peak mass greater than fifty times the particle mass, $m_{\rm p}$. Similarly, $V_{\rm peak}$ is the peak value of the maximum circular velocity, $V_{\rm max}$, a (sub-)halo obtains over the course of its history, and  $z_{\rm acc}$ is (roughly) the redshift at which a sub-halo is first accreted into a more massive host halo.  A detailed description of how each of these quantities is calculated can be found in Appendix \ref{sec:halo_properties}.

\begin{figure*}
    \includegraphics[width=\textwidth]{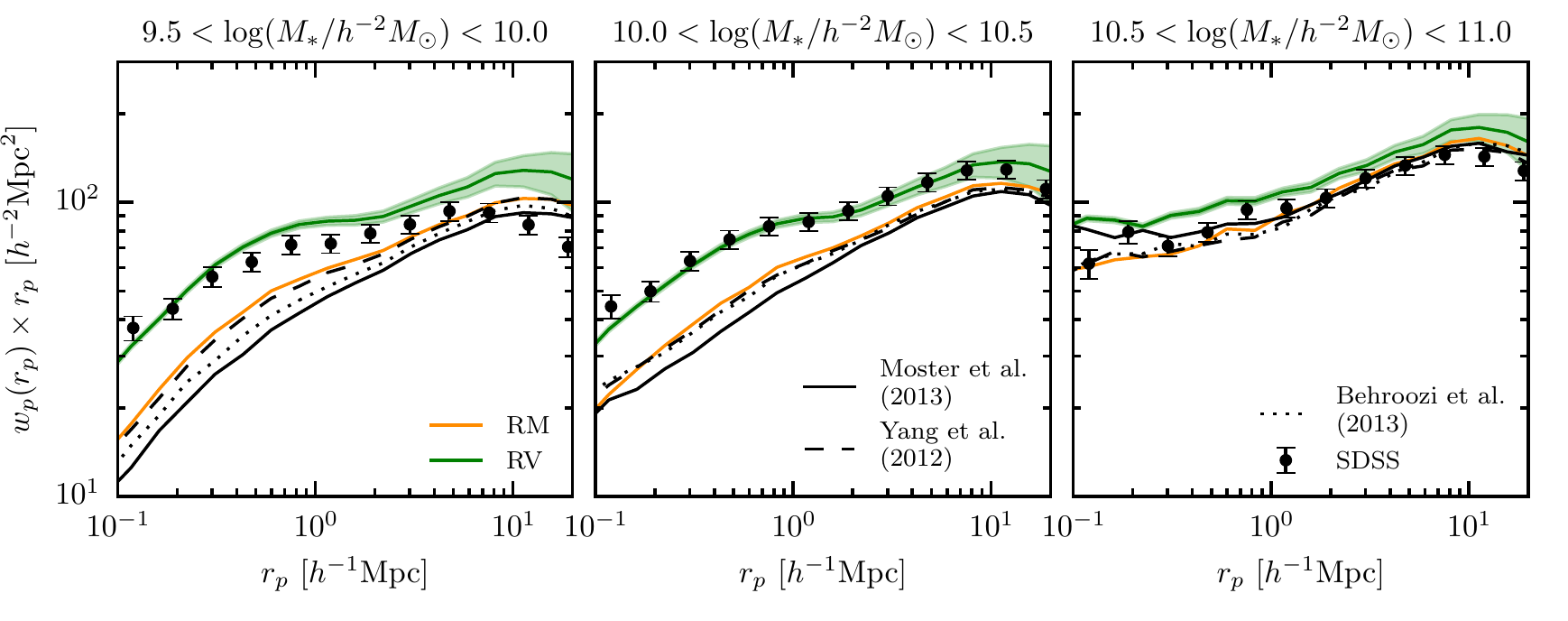}
    \caption{The projected correlation function, $w_p$, is plotted in three stellar mass bins for five SHAM models: RM (solid orange), RV (solid green), M13 (solid black), Y12 (dashed), and B13 (dotted).  As an example, we show the $\pm 1\sigma$ error on the model prediction for the RV model as the shaded green region.  For comparison, we plot the projected correlation function of galaxies in SDSS as measured by \citet{Yang:2012ew} as points with error bars.}
    \label{fig:wp_comparison}
\end{figure*}

\subsection{Populating Simulations}
\label{sec:pop}

We build mock galaxy catalogues using the models and simulation described in the previous sections.  For the rank order SHAM models, RM and RV, we use the triple piece-wise Schechter function fit to the stellar mass function, $\phi_{\rm SDSS}$, from \citet[][LW09 hereafter]{Li:2009kh}.  This stellar mass function is based on a complete sample of galaxies from SDSS \citep{York:2000gn} and assumes a universal \citet{Chabrier:2003ki} initial mass function (IMF).  We integrate the stellar mass function to get the cumulative stellar mass function:
\begin{equation}
N(>M_*) = V\int_{M_*}^{\infty} \phi_{\rm SDSS}(M_*^{\prime}) \mathrm{d}M_*^{\prime}
\end{equation}
where $V$ is the volume of the simulation being populated.  We then normalize by the total number of galaxies above the threshold, $N_{\rm lim} = N(>10^{9.5}~h^{-2}M_{\odot})$, to get $F(M_*) = N(>M_*)/N_{\rm lim}$, the cumulative probability distribution of galaxies as a function of stellar mass.  We then draw from this distribution $N_{\rm lim}$ times using the inverse transform sampling method.  In this way, each sampling is a Monte Carlo (MC) realization of the stellar mass function in the simulation volume.  (Sub-)haloes are then populated by matching rank orders between stellar mass and (sub-)halo mass such that the most massive (sub-)haloes receive the most massive galaxies.

The parametrized SMHM models are populated in a different manner.  The population of the simulation is a MC realization which begins by calculating the median stellar mass for each (sub-)halo and adding random scatter:
\begin{align}
\label{eq:mstar_assign}
\log[M_*(M_{\rm peak}, a_{\rm acc})] =& \log[\langle M_* | M_{\rm peak}\rangle(a_{\rm acc})] \\ \nonumber
&+ \mathcal{N}(0,\sigma_{\log(M_*)})
\end{align}
where $\mathcal{N}(0,\sigma_{\log(M_*)}$ is a random variable drawn from a normal distribution with mean 0 and a fixed log-scatter given by $\sigma_{\log(M_*)}$.  We use $\mathcal{M} = M_{\rm peak}$ for all the models, but we confirm that using $M_{\rm acc}$ instead does not significantly affect our results.

This step is slightly complicated by the fact that each model's parameters were tuned using different halo mass definitions, $M_{\Delta_{\rm halo}}$. The M13 model uses $M_{200\rm c}$,  Y12 uses $M_{180\rm m}$, and B13 uses $M_{360\rm m}$\footnote{The subscript ``m'' indicates the over-density with respect to the mean density of the Universe, while ``c'' is with respect to the critical density.}.  As described in the previous section, our halo catalogue defines haloes as SO with masses given by $M_{360\rm m}$.  We use $M_{\rm peak}$ directly for the B13 model, and for the M13 and Y12 models we convert $M_{\rm peak}$ from $M_{360\rm m}$ masses to the appropriate version for each model using a fitting function, $f_{\rm conv}$ \citep[see appendix C in][]{Hu:2003dr}.  In the conversion we assume all haloes are fit by an NFW profile \citep{Navarro:1997if} and use halo concentrations measured by {\tt ROCKSTAR} at the time $M_{\rm peak}$ is achieved, $c_{\rm peak}$.
\begin{equation}
M_{\rm peak}^{\prime} = f_{\rm conv}(M_{\rm peak}, c_{\rm peak}, \Delta_{\rm halo})
\end{equation}

Furthermore, each of the evolving models was tuned using observations based on different methods of determining stellar mass.  We apply a simple set of conversions in order to homogenize the stellar masses to a single system as described in Appendix \ref{appendix:stellar_mass_conv}.   We show the stellar mass function for a mock based on each model in Fig. \ref{fig:phi}.  The stellar mass functions in each mock vary by up to $\sim 0.2$ dex from our fiducial stellar mass function used to create the RM and RV mocks.  This is the result of each model being tuned with differing:
\begin{enumerate}
\item stellar mass functions, $\phi(M_*,z)$
\item halo mass definitions/(sub-)halo finder
\item cosmologies
\item simulations which are subject to cosmic variance
\end{enumerate}
While in principle each model could be re-fit using the same stellar mass function(s) and simulation, that is beyond the scope of this paper.  We check that simple alterations to the parameters of each model which bring each into better agreement with the LW09 stellar mass function do not have a significant effect on the clustering signal predictions presented in \S \ref{sec:clustering}.  Given this, and the complexity involved in re-fitting each model, we use the parameters as they are quoted in each paper (and reproduced in table 2, with only minor alterations as discussed in the sections above).

\section{Clustering}
\label{sec:clustering}

In this section we present the galaxy clustering predictions of each SHAM model described in the previous section.  We calculate the projected correlation function for each model, defined as:
\begin{equation}
w_p(r_p) = 2\int_0^{r_{\parallel \rm max}}\xi(r_{\parallel},r_p)\mathrm{d}r_{\parallel}
\end{equation}
where $r_{\parallel}$ is the separation parallel to the line-of-sight (LOS), $r_{p}$ the separation perpendicular to the LOS, and $\xi$ is the two-point correlation function.  We set $r_{\parallel \rm max} = 40~{h^{-1}\rm Mpc}$ to mimic equivalent measurements made using SDSS galaxies \citep{Yang:2012ew}.  The choice of $r_{\parallel \rm max}$ is a balance between minimizing the contribution of redshift space distortions to the measurement, and maximizing the signal to noise in the measurement \citep{Padmanabhan:2007kn, vandenBosch:2013jf}.

We assume the ``distant observer'' approximation when calculating $w_{p}$.  We approximate the LOS direction to a galaxy as $\hat{z}$, and take the plane-of-the-sky to be coincident with the $x-y$ plane of the simulation box.  This assumption holds when the distance between the ``observer'' and galaxies is much larger than the maximum separation between galaxies which enters the calculation for a given statistic.  Since we measure $w_{p}$ only out to $20$ Mpc, this is a safe approximation.

Any residual effects should be minimal because we also place galaxies in each mock in ``redshift space''.  Galaxies are assigned the peculiar velocity of the (sub-)halo which they occupy, $\vec{v}_{\rm pec}$.  The cosmological redshift, $z_{\rm cosmo}$, is calculated using the z-position of the galaxy in the mock.  The ``observed'' distance to a galaxy, $z^{\prime}$, is then:
\begin{equation}
\mathrm{z}^{\prime} = [z_{\rm cosmo} + \frac{v_{\rm pec}}{c} \,(1+z_{\rm cosmo}) ] \, \frac{c}{H_0}
\end{equation}
where we additionally account for the periodic boundary conditions of the simulation.  We estimate jackknife errors on $w_p$ by sub-dividing the simulation box into 27 equal cubic sub-volumes.

\begin{figure}
    \includegraphics[width=\columnwidth]{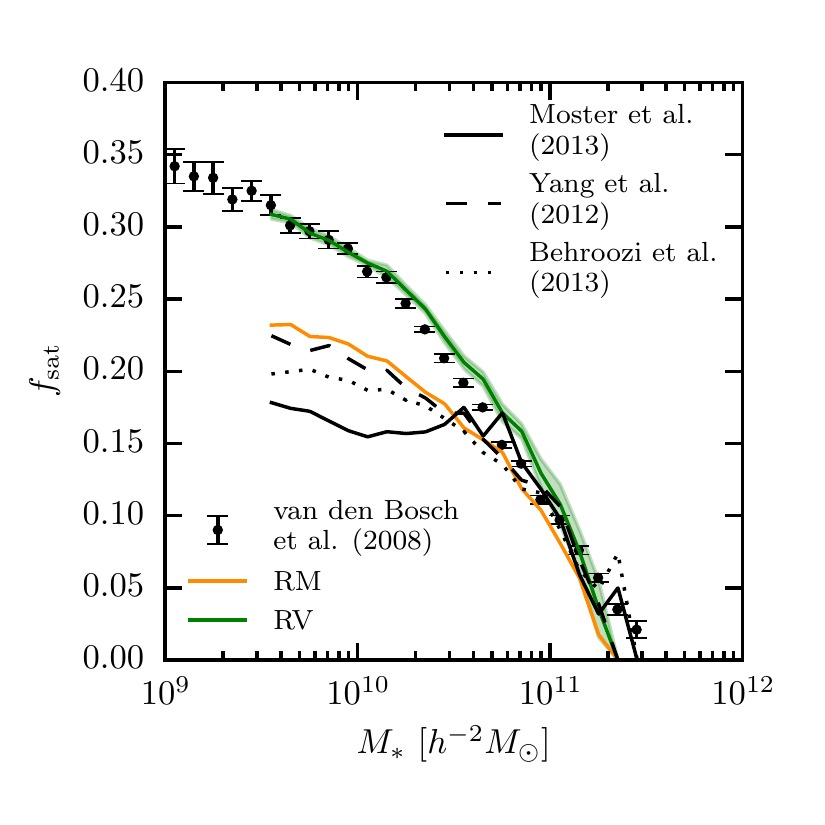}
    \caption{The satellite fraction as a function of stellar mass is plotted for the five SHAM models as lines.  The models are RM (solid orange), RV (solid green with the shaded region showing $\pm 1\sigma$ error on the model prediction), M13 (solid black), Y12 (dashed), and B13 (dotted).  For comparison, we show the satellite fraction as inferred by \citet{vandenBosch:2008fv} using a galaxy group catalogue of SDSS galaxies.}
    \label{fig:f_sat_mstar}
\end{figure}

The projected correlation functions for each model calculated in three stellar mass bins as described above are shown in Fig. \ref{fig:wp_comparison}.  For comparison, we show the corresponding measurement made on a galaxy sample in SDSS \citep[``Mass-limit'' sample in][]{Yang:2012ew}.  The difference between the clustering signal predictions of each SHAM model and observations is striking.  Each of the halo mass-based models\footnote{As a short-hand, we will frequently refer to the RM, M13, B13, and Y12 models as ``mass-based models''.} severely under-predicts the clustering signal in the two least massive bins (left two panels), while the RV model, based on $V_{\rm peak}$, if anything, slightly over-predicts the signal.  On the other hand, all models predict a clustering signal for the most massive stellar mass bin (right panel) that is roughly consistent with observations.

The success of the RV model is consistent with findings by \cite{Reddick:2013gi} who show that SHAM based on $V_{\rm peak}$ most closely matches galaxy clustering observations in SDSS, while all other halo properties under-predict the signal.  We confirm that result here; furthermore, we show that more complicated evolving SMHM models based on $M_{\rm peak}$ do not alleviate the problem.  In fact, evolution seems to exacerbate the clustering decrement of $M_{\rm peak}$ based SHAM as evidenced by the RM model producing the strongest clustering signal amongst the mass-based models.  The remainder of this paper examines why these models fail to produce sufficiently strong clustering, and why the RV model (and similar models) succeeds. In
addition, we explore possible solutions to ``save'' mass-based SHAM models (under the assumption these models are worth saving).

\section{The $V_{\rm peak}$ Miracle}
\label{sec:vpeak_miracle}

In this section, we examine the origin of the success of $V_{\rm peak}$ based SHAM models.  By comparing the differences in $V_{\rm peak}$ and $M_{\rm peak}$ based models, we identify the culprit(s) in the failure of the mass-based models to match observational clustering measurements.

\subsection{The Satellite Fraction}

In the mass-based models the clustering signal is most severely under-predicted at small scales $(<1~h^{-1}{\rm Mpc})$.  This is a strong indication that the culprit is satellite galaxies--or more precisely, a lack of satellite galaxies in the mass-based SHAM models.  With this in mind, we examine the satellite fraction, $f_{\rm sat}$, in each model in Fig. \ref{fig:f_sat_mstar}.  The RV model results in the largest number of satellites, approaching $\sim30\%$ at $10^{9.5}~h^{-2}M_{\odot}$, relative to the other models which remain below  $\sim24\%$. Furthermore, the RM model results in a satellite fraction in remarkable agreement with $f_{\rm sat}$ derived from a SDSS galaxy group catalogue \citep{vandenBosch:2008fv}.  This suggests that the satellite fraction in RV is quite realistic, while it is under-predicted in the mass-based models. 

Comparing the clustering signals in Fig. \ref{fig:wp_comparison} with $f_{\rm sat}$ in Fig. \ref{fig:f_sat_mstar} shows that there is a nearly one-to-one correspondence between the clustering strength on small scales and $f_{\rm sat}$ in each model.  Furthermore, $f_{\rm sat}$ between the models does not diverge until approximately $5 \times 10^{10}h^{-2}M_{\odot}$, above which the models roughly agree with the observed clustering signal.  Considering this, it may be more appropriate to restate the under-prediction of galaxy clustering in these models as an under-production of satellite galaxies \citep[][]{Lehmann:2017fy}.

\subsection{Why is $f_{\rm sat}$ larger in $V_{\rm peak}$ SHAM?}

\begin{figure}
    \includegraphics[width=\columnwidth]{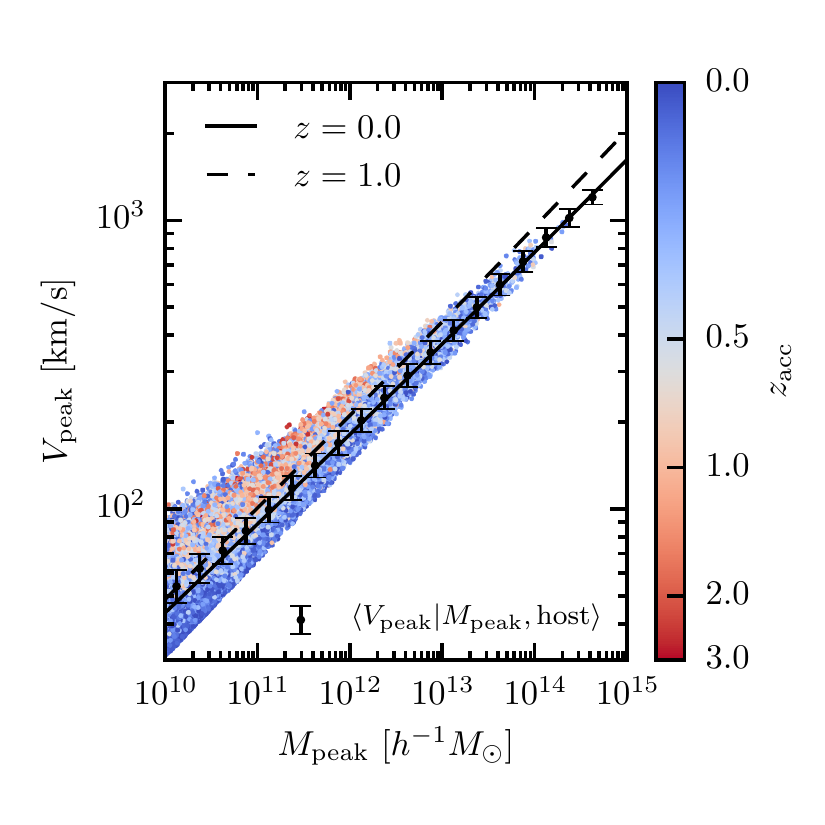}
    \caption{The relation between $M_{\rm peak}$ and $V_{\rm peak}$ is plotted for all sub-haloes in the Bolshoi simulation at z=0, colour-coded according to the redshift at the time of accretion, $z_{\rm acc}$.  For host-haloes, the median $V_{\rm peak}$ in 0.25 dex bins of $M_{\rm peak}$ is plotted as black points with error bars indicating the log-normal scatter.  The solid line shows the median $V_{\rm peak}$ relation in eq. \ref{eq:vpeak} at z=0.  The dashed line shows the same relation at z=1.  Sub-haloes have systematically higher values of $V_{\rm peak}$ than host-haloes at fixed $M_{\rm peak}$.}
    \label{fig:mpeak_vpeak}
\end{figure}

As can be seen in Fig. \ref{fig:mpeak_vpeak}, $V_{\rm peak}$ and $M_{\rm peak}$ are highly correlated properties of haloes with small scatter, $\sigma_{\log(V_{\rm peak})}\sim 0.05$, at fixed $M_{\rm peak}$ for host-haloes.  Additionally, sub-haloes have larger $V_{\rm peak}$ on average than host-haloes of equal $M_{\rm peak}$.  This difference between host-haloes and sub-haloes is a result of the correlation between halo formation history and concentration.  

Assuming an NFW profile, $V_{\rm max}$ is an increasing function of the halo mass and concentration, $c$:
\begin{equation}
V_{\rm max} = 0.465V_{\rm vir}\sqrt{\frac{c}{f(c)}}
\end{equation}
where,
\begin{align}
V_{\rm vir} = ~&159.43~{\rm km/s}\left(\frac{M_{\rm vir}}{10^{12}h^{-1}M_{\odot}}\right)^{1/3}\left[\frac{H(z)}{H_0}\right]^{1/3} \\ \nonumber
 &\times \left[\frac{\Delta_{\rm vir}(z)}{178\Omega^{-1}_m}\right]^{1/6}
\end{align}
and,
\begin{equation}
f(x) = \ln(1+x) - \frac{x}{1+x}
\end{equation}
Furthermore, it is a robust prediction of simulations of $\Lambda$CDM structure formation that halo concentration correlates with halo formation time, $\langle z_f \rangle = g(M_{\rm vir}, c)$ \citep[e.g.][]{Wechsler:2002kh}.  Finally, because $M_{\rm peak}$ occurs before accretion for sub-haloes and near z=0 for host-haloes, sub-haloes generally have earlier formation times at fixed $M_{\rm peak}$.   

Using the concentration-mass-redshift relation from \citet{Maccio:2009bs}, we formulate a description for the median $M_{\rm peak}-V_{\rm peak}$ relation given by:
\begin{equation}
\label{eq:vpeak}
\langle V_{\rm peak} | M_{\rm peak}\rangle_{\rm med}(z) = 1.1\times \langle V_{\rm max}|M_{\rm peak}\rangle_{\rm med}(z)
\end{equation}
The factor of $1.1$ accounts for the fact that the average peak maximum circular velocity is $\sim 10\%$ higher than $V_{\rm max}$ \citep{Behroozi:2014bg}.  We show this relation at $z=0$ (solid line) and at $z=1.0$ (dashed line) in Fig. \ref{fig:mpeak_vpeak}.  At fixed $M_{\rm peak}$, $\langle V_{\rm peak}\rangle_{\rm med}$ is larger at higher redshifts. 

It is the correlated scatter in $V_{\rm peak}$ at fixed $M_{\rm peak}$ at $z=0$ which is responsible for the difference in satellite fractions between the RV and RM models.  By populating (sub-)haloes by their rank on $V_{\rm peak}$, more sub-haloes will be populated than would have been had $M_{\rm peak}$ instead been utilized for any given stellar mass threshold.  This same reasoning carries over to the other mass-based models.  

\begin{figure}
    \includegraphics[width=\columnwidth]{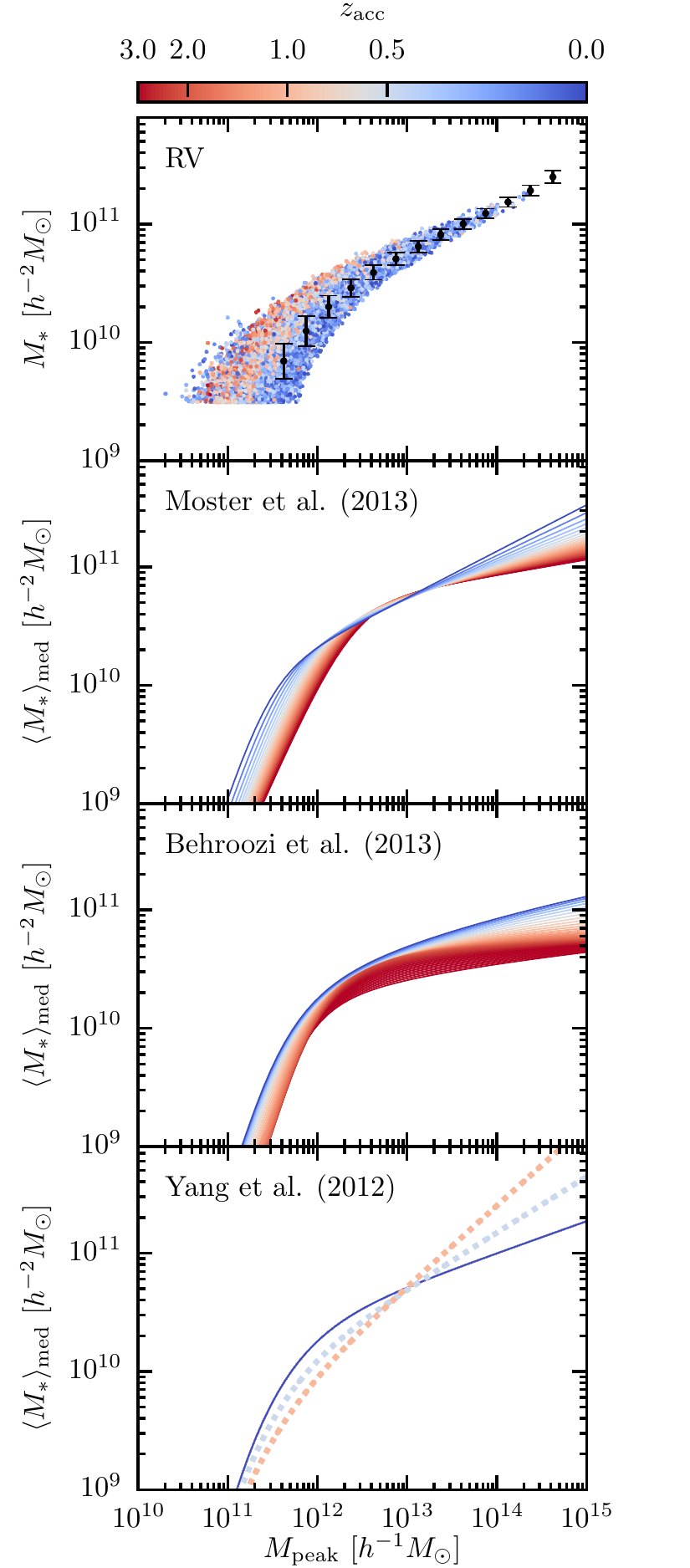}
    \caption{The SMHM relation in the RV model and the three evolving SHAM models.  From top to bottom, the panels show RV, M13, B13, and Y12.  For each model, the SMHM relation is shown at z=0 as a function of the redshift of accretion, $z_{\rm acc}$, for satellites.  The top panel shows the median stellar mass and scatter of central galaxies as points with error bars with satellites shown as colour-coded points.  For the remaining panels, the analytic functions are shown. For these models, the relation for centrals is shown with $z_{\rm acc}$ set to $0$.  The dotted lines in the bottom panel give the relation for central galaxies at $z=1$ and $z=0.5$ as an example of the relation satellites that accreted at that time would follow if there were no evolution after accretion in the Y12 model (i.e., if the $c$ parameter
were equal to $0.0$ rather than $1.0$).}
    \label{fig:SMHM_comparison}
\end{figure}

\subsection{The Evolving Relation between $M_*$ and $M_{\rm halo}$}
\label{sec:rv_sm_evolv}

Here we compare the RV model to the other models by examining the relation between stellar mass and halo mass that arises in each model, in the RV model by correlation between $V_{\rm peak}$ and $M_{\rm peak}$, and in the other models by design.  In the top panel of Fig. \ref{fig:SMHM_comparison}, we examine the relation between $M_*$ and $M_{\rm peak}$ in the RV model.  The median stellar mass mass of central galaxies is shown as points with error bars indicating the log-normal scatter at fixed halo mass.  The colour-coded points show the stellar mass of satellites.  Beyond the ubiquitous broken power-law shape of the SMHM relation, sub-haloes ($z_{\rm acc}>0$) host more massive galaxies than host-haloes of equivalent $M_{\rm peak}$, with earlier accreted satellites having the largest masses.  This is precisely the expectation for satellites given the correlation between $V_{\rm peak}$ and formation time shown in the previous section.

We show the SMHM relation for the M13, B13, and Y12 models in the lower panels of Fig. \ref{fig:SMHM_comparison}; however, instead of showing individual satellite galaxies, we show the median relation as a function of $z_{\rm acc}$ for simplicity and ease of interpretation.  To the degree that satellites follow a different relation than centrals, the mass-based models predict the qualitative opposite trend as the RV model; sub-haloes host {\em less} massive galaxies than host-haloes of equivalent $M_{\rm peak}$ (at least for haloes less massive than $\sim 10^{12.5} ~ h^{-1} M_{\odot}$).  In the M13 and B13 models, this comes about because the SMHM relation for satellites as a function of $z_{\rm acc}$ serves as a fossil record of the SMHM relation for centrals at higher redshift, and \citet{Moster:2013ab} and \citet{Behroozi:2013fg} find that the SMHM relation must evolve such that $M_*$ at fixed $M_{\rm peak}$ increases over cosmic time. Note that this holds because it is assumed that satellites cease to grow after being accreted. 

On the other hand, in the Y12 model there is no difference in $\langle M_*\rangle_{\rm med}$ between satellites and centrals at fixed $M_{\rm peak}$.  The Y12 model is different from M13 and B13 in that satellites are allowed to grow (or lose stellar mass) after $z_{\rm acc}$.  \citet{Yang:2012ew} find that in the Y12 model post-accretion evolution happens very efficiently--satellites achieve a final stellar mass $> 90\%$ that of centrals by $z=0$.  For simplicity, in this paper we have assumed that satellites obtain the same stellar mass as centrals ($c$ parameter set to 1.0 in Y12), consistent with the uncertainty found by \citet{Yang:2012ew}.  Therefore, all of the lines fall on top of one another in the bottom panel of Fig. \ref{fig:SMHM_comparison}.  For comparison, the dotted lines show the Y12 SMHM relation for centrals at $z=0.5$ and $1.0$. If there were no evolution in the stellar mass of satellite after accretion ($c$ set to $0.0$), then this would reflect the
stellar masses of satellite galaxies accreted at $z_{\rm acc} = 0.5$
and $1.0$, respectively. The sense of evolution in this case is similar to the M13 and B13 models below $\sim 10^{13}~h^{-1}M_{\odot}$.

This model for satellite growth in Y12 is the reason why it predicts a slightly larger $f_{\rm sat}$ than the M13 and B13 models (as shown in Fig \ref{fig:f_sat_mstar}).  Because the sub-halo fraction increases as $M_{\rm peak}$ decreases (see \S \ref{sec:orphans} and Fig. \ref{fig:f_sub}), any model which boosts $M_*$ in sub-haloes, will result in a larger fraction of satellites for any given stellar mass threshold.  The same reasoning holds for the RM model where centrals and satellites follow the same SMHM relation.  However, this effect in the Y12 and RM models alleviates the clustering signal decrement only slightly compared to M13 and B13.  In the absence of post-accretion evolution in $M_*$ of satellites in Y12 ($c$ parameter set to 0), the Y12 model appears similar to the M13 and B13 models where satellites are less massive than centrals at z=0.0 (below $\sim 10^{12.5} ~ h^{-1} M_{\odot}$), and the galaxy clustering signal is further weakened on small scales.  We examine post-accretion evolution of satellites in more detail in \S \ref{sec:sat_growth}.

We draw two conclusions from the comparison between the SMHM relation in each of the SHAM models: First, satellites are more massive than central galaxies at fixed $M_{\rm peak}$ in the RV model, and second, the evolving models tend to require satellites be less massive than centrals at fixed $M_{\rm peak}$ in the absence of post-accretion growth.  At first glance, the evolution in the SMHM relation in the evolving models seems incompatible with RV.  It could have been the case that evolving SMHM models predicted evolution in the opposite direction, and thus produced more massive satellites and stronger clustering on small scales.  This would naturally provide a physical explanation for an RV-like model.

\section{A Catch-22}
\label{sec:not_such_a_miracle}

In this section we point out some shortcomings in $V_{\rm peak}$-based SHAM when $V_{\rm peak}$ is taken to be the physical property that determines stellar mass, and the tension in fitting both observed galaxy clustering measurements at $z\sim 0$ and galaxy growth histories {\em simultaneously}.

\subsection{Galaxy Growth Histories}

\begin{figure}
    \includegraphics[width=\columnwidth]{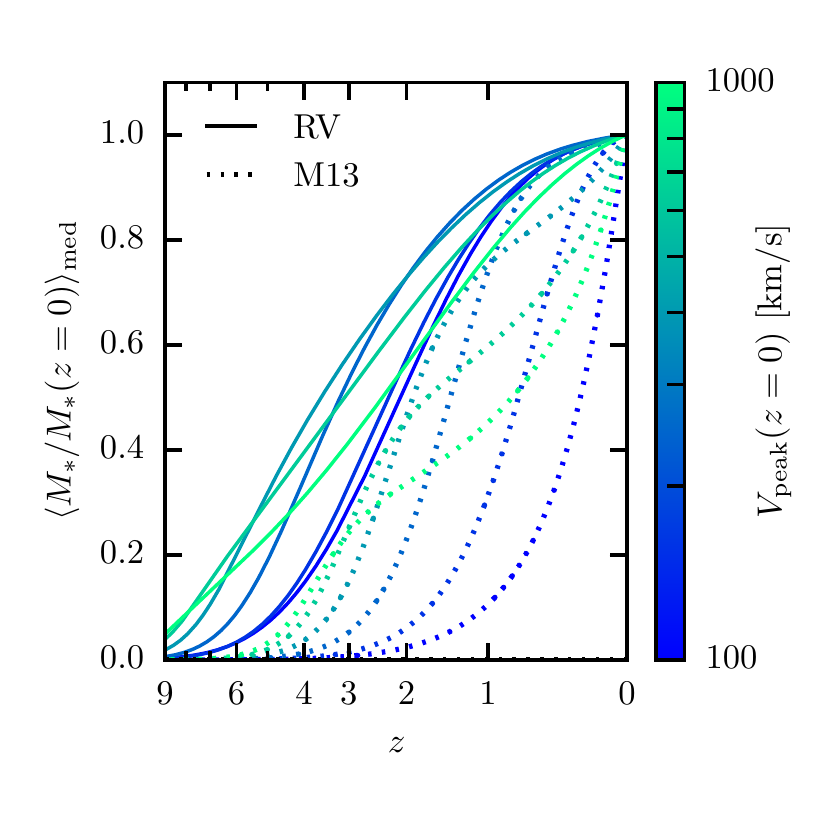}
    \caption{The median stellar mass as a function of redshift divided by the mass at z=0 for haloes of various values of $V_{\rm peak}$ at z=0 (coloured-coded lines).  The prediction from the non-evolving RV model is shown as solid lines.  The prediction of the evolving M13 model is shown as dotted lines.  $M_{\rm peak}(z=0)$ was converted to $V_{\rm peak}$ using a fitting function in order to directly compare the RV and M13 models.  The implied average stellar mass growth histories of haloes are drastically different between the two models.  The RV model predicts {\rm much} earlier stellar mass growth relative to the M13 model.}
    \label{fig:mstar_of_z_rm}
\end{figure}

\begin{figure}
    \includegraphics[width=\columnwidth]{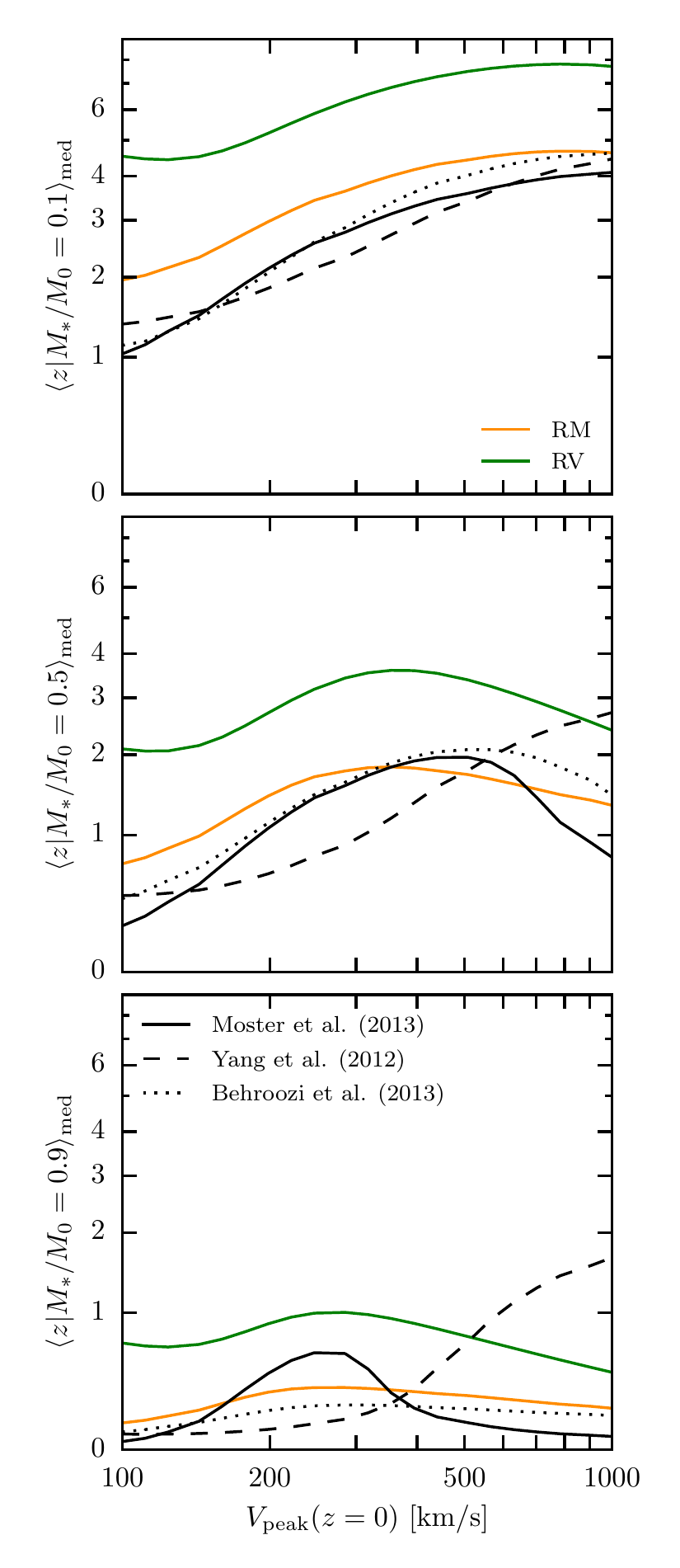}
    \caption{The median redshift at which a central galaxy reaches a fraction, $f$, of its $z=0$ stellar mass as a function $V_{\rm peak}$ at $z=0$. The three panels are for $f=0.1, 0.5$, and $0.9$ from top to bottom, respectively.  The lines are for the RM (orange), RV (green), M13 (solid black), Y12 (dashed), and B13 (dotted).  The RV model consistently forms a larger fraction of galaxies' mass at higher redshift relative to the mass-based models.}
    \label{fig:mstar_zform}
\end{figure}

Given that the RV and mass-based models make different assumptions for the SMHM relation, especially regarding satellites, in hindsight it is not surprising that each model predicts different galaxy clustering signals and satellite abundances.   Considering the RV model's success in fitting clustering observations and the mass-based models' failure (see \S \ref{sec:clustering}), it appears that the RV model should be favoured as the more physical.  However, in this section we show that the RV model, if taken at face value, implies galaxy growth histories that are incompatible with both the other models and observations.

\begin{figure*}
    \includegraphics[width=\textwidth]{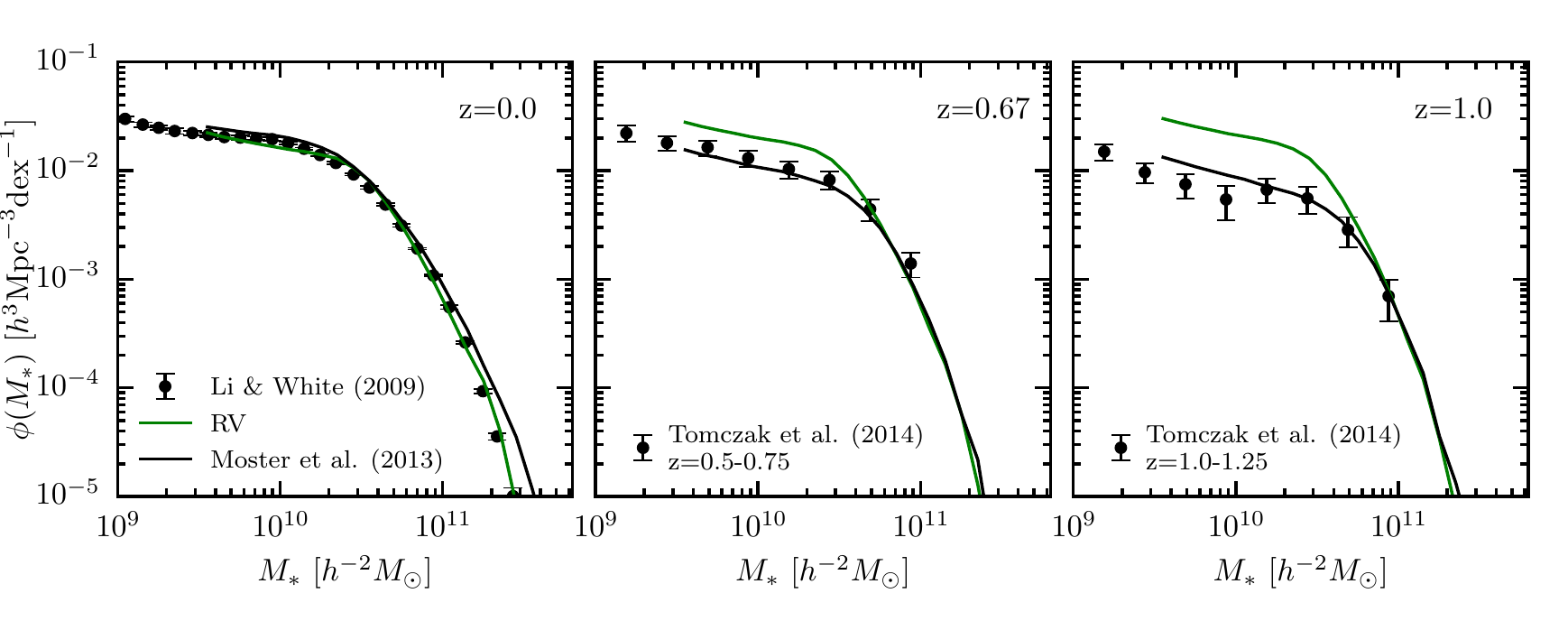}
    \caption{The evolution of the stellar mass function is shown for the M13 model (black lines) as well as the implied evolution of the RV model (magenta lines). For comparison, we show observational measurements of the stellar mass function from  \citet{Li:2009kh} ($z=0$), and \citet{Tomczak2014:db} ($z>0$).}
    \label{fig:stellar_mass_function_evolution}
\end{figure*}

In the RV model, (sub-)haloes with the same $V_{\rm peak}$ are assigned the same $M_*$ regardless of the redshift at which $V_{\rm peak}$ is achieved. For example, a host-halo at $z\sim 0$, with a particular value of $V_{\rm peak}$ will be assigned precisely the same stellar mass (aside from any possible scatter in the model) as all sub-haloes with the same value of $V_{\rm peak}$. However, host-haloes achieve $V_{\rm peak}$ at a time very close to $z=0$, whereas $V_{\rm peak}$ for sub-haloes is achieved prior to accretion into the parent halo, often at considerably higher redshift \citep{Reddick:2013gi, Behroozi:2014bg, Mao:2017ww}. SHAM models of this kind cannot have evolution because they implicitly assume that the $M_*-V_{\rm peak}$ (SMVP) relation is identical at all redshifts. In other words, taken at face value, the RV model implies a SMVP relation that is independent of redshift\footnote{We speculate on evolution in the SMVP in \S \ref{sec:vpeak_or_mpeak}.}. In this subsection, we examine the implied growth histories of galaxies in such a model with a universal, non-evolving, SMVP relation. 

In order to infer smooth growth histories below $10^{9.5}~h^{-2}M_{\odot}$, we fit the SMVP relation in the RV model with a function of the form:
\begin{equation}
\label{eq:smvp_relation}
\langle M_* | V_{\rm peak} \rangle_{\rm med} = 2M_0\left(\frac{V_{\rm peak}}{V_0}\right)\left[\left(\frac{V_{\rm peak}}{V_0}\right)^{\alpha}+\left(\frac{V_{\rm peak}}{V_0}\right)^{\beta}\right]^{-1}
\end{equation}
We perform a non-linear least squares fit to the median stellar mass in 0.025 dex $V_{\rm peak}$ bins.  We find an excellent fit is provided by: $\log(M_0/h^{-2}M_{\odot})=9.95 \pm 0.01$, $\log(V_0/{\rm km}\, {\rm s}^{-1})=2.177 \pm 0.005$, $\alpha=-5.9 \pm 0.1$, and $\beta=-0.25 \pm 0.02$.  We perform a similar fit to the SMHM relation in the RM model.

In addition to the form of the SMVP relation, the average growth history of galaxies in RV is dependent on the potential well growth history (PWGH) of haloes, $\langle V_{\rm peak}| V_{\rm peak}(z=0) \rangle_{\rm med}(z)$.  Using the PWGHs from \citet{Bosch:2014cu}, in Fig. \ref{fig:mstar_of_z_rm} we show the implied growth history of galaxies in the RV model for host-haloes with different $V_{\rm peak}$ at $z=0$.  For comparison, we also show stellar mass growth histories from M13, which instead depend on the average mass accretion histories (MAHs) of haloes and the explicitly parametrized evolution of the SMHM relation.  Note that we have converted $M_{\rm peak}(z=0)$ to $V_{\rm peak}(z=0)$ using eq. \ref{eq:vpeak} to place these histories on the same figure.  

Fig. \ref{fig:mstar_of_z_rm} shows that, in the RV model, galaxy growth is largely self-similar--at any given redshift, galaxies of every mass grow at similar rates.  Conversely, in the M13 model, galaxy growth is much more dependent on halo mass--high mass galaxies form early, growing slowly at low redshift, while low mass galaxies form late.  Slow late time growth of massive galaxies is necessary to reproduce the observed prevalence of quiescent galaxies as mass increases \citep[e.g.][]{vandenBosch:2008fv, Wetzel:2012lk}.  This is related to the downsizing phenomenon \citep{Neistein:2006fl}, wherein star-formation shifts to less massive galaxies at low redshifts \citep[e.g.][]{Kodama:2004gb, Jimenez:2005gj, Juneau:2005ig, Bell:2005hs, Bundy:2006em}.  

It is the same story when comparing RV to any of the other mass based models.  In Fig. \ref{fig:mstar_zform}, we show the median redshift
at which galaxies formed a fraction, $f$, of their $z=0$ stellar mass (this is another way of examining the information in Fig. \ref{fig:mstar_of_z_rm}).  The top panel shows the median redshift at which galaxies formed 10\% of their mass.  The middle and bottom panels show the equivalent redshifts but for 50\% and 90\%, respectively.  The primary conclusion to draw from this is that the mass-based SHAM models generically predict much later growth compared to RV with a non-evolving SMVP.  Interestingly, even the RM model, where the SMHM relation does not evolve, results in galaxy growth that is much closer to the evolving models than to the predictions for the RV model.  This is a consequence of the fact that haloes form their potential wells early, primarily adding mass in the outskirts at late times \citep{Bosch:2014cu}.

Finally, the effect of the dramatically different growth history in the RV model can be seen in the evolution of the stellar mass function.  In Fig. \ref{fig:stellar_mass_function_evolution} we show the predicted stellar mass function, $\phi(M_*,z)$, in the RV model. For comparison, we also show the predictions for the M13 model, which was tuned to reproduce this evolution in the stellar mass function. It is clear that the RV model predicts galaxy abundances that are too large at high redshift, particularly below the knee in the mass function.

\subsection{$V_{\rm peak}$ or $M_{\rm peak}$?}
\label{sec:vpeak_or_mpeak}

Given that the M13, Y12, and B13 models were fit to the stellar mass function at various redshifts and reproduce the cosmic star-formation density in the Universe (among other observables), it is difficult to imagine how to make a model like RV consistent with these same observational constraints.  While not explored here, it is likely that a more complicated $V_{\rm peak}$-based SHAM model could produce realistic galaxy growth histories, but we have shown that such a model would require evolution in the SMVP relation. We see two distinct options for an evolving $V_{\rm peak}$-based\footnote{or related $V_{\rm max}$ derived quantities} SHAM model approach: 1.) perform rank order SHAM using $V_{\rm peak}$ at each redshift, or 2.) parametrize the evolution of the SMVP relation and assume a model for post-accretion satellite evolution.  The former will reproduce the observed $\phi(M_*, z)$ by design and the $z\sim 0$ observed galaxy clustering signal, but this approach implicitly relies on satellites adjusting their stellar mass such that they remain on the evolving SMVP relation.  The latter approach will be subject to the same problem as the evolving mass-based SHAM models where satellites tend to be less massive than centrals. The SMVP relation will be required to decrease as redshift increases. As a result, satellites will be less massive at $z_{\rm acc}$, which, as we have shown, generally results in a decreased small scale clustering signal. We leave a formulation of an evolving $V_{\rm peak}$-based SHAM model to future work.

Given the lack of a coherent picture for galaxy growth in a $V_{\rm peak}$ SHAM model, $V_{\rm peak}$ is not clearly preferable over $M_{\rm peak}$ as a physically motivated SHAM parameter \citep[e.g. see discussion in][]{Lehmann:2017fy}.  Furthermore, apart from arguments related to fitting $w_p(r_p)$ and $\phi(M_*,z)$, there are other difficulties that must be overcome if $V_{\rm peak}$ is to be argued as more fundamental in driving the stellar mass mapping into (sub-)haloes.  For any given (sub-)halo, $V_{\rm peak}$ is generally set during major mergers \citep[1:5 or larger,][]{Behroozi:2014bg}. If stellar mass is tightly correlated with $V_{\rm peak}$, this could imply bursty star-formation closely tied to major mergers which is not favoured by observations \citep[e.g. star formation histories,][]{Diemer:2017gj}. However, we note that this may not be a problem with other closely related quantities like $V_{\rm max}$ at the time $M_{\rm peak}$ is achieved, or $V_{\rm relax}$ \citep{ChavesMontero:2015dc}.  On the other hand, $M_{\rm peak}$ remains a theoretically attractive quantity that should be tightly coupled to stellar mass. Halo mass is also set later than $V_{\rm max}$ and therefore stellar mass growth suffers from less of a downsizing effect.  Peak halo mass is plausibly a good indicator of the amount of gas that has been accreted onto a host-halo and in principle available to a galaxy to form stars over its history \citep{White:1978uk, Blumenthal:1986ie}. Nevertheless, it is possible that feedback processes that modulate star-formation efficiency happen to correlate with $V_{\rm peak}$, or $V_{\rm peak}$ correlates with environmental parameters that affect star-formation.  It remains possible that the galaxy-halo connection in the RV model at $z\sim 0$ is closer to reality than RM or the evolving mass-based models; however, it is not clear how this relation comes about, especially with respect to satellites.

As it stands, there is a tension between fitting galaxy clustering on small scales while simultaneously reproducing the build-up of stellar mass in the Universe using SHAM techniques.  This tension presents a ``clustering crisis'' for SHAM.  Mass-based models which are fit to reproduce the evolution of the stellar mass function universally under-predict galaxy clustering on small scales.  $V_{\rm peak}$-based models that fit galaxy clustering at $z\sim 0$ are either unable to fit the evolution of the stellar mass function or it remains to be seen how to add in evolution in a consistent manner.     

Our exploration of SHAM models suggests multiple ways the mass-based models\footnote{We leave an exploration of modifications to $V_{\rm peak}$-based SHAM models to future work.} could be altered in order to alleviate this crisis. We have identified too few satellites as the primary culprit for the clustering deficiency in mass-based SHAM models.  One avenue to address this problem is to simply increase the number of satellites in the mass-based models as many studies have found to be necessary.  In \S \ref{sec:orphans}, we examine the plausibility of missing sub-haloes, and thus satellites, in our implementation of the mass-based SHAM models.  A significant population of missing sub-haloes would be an indication that so called ``orphan'' galaxies play an important role in solving the clustering crisis.  Apart from orphans, we also consider two other physically motivated modifications to increase the satellite fraction: post-accretion satellite growth in \S \ref{sec:sat_growth} and assembly bias effects in \S \ref{sec:assem_bias}.   

\section{Orphan Galaxies}
\label{sec:orphans}

\begin{figure}
    \includegraphics[width=\columnwidth]{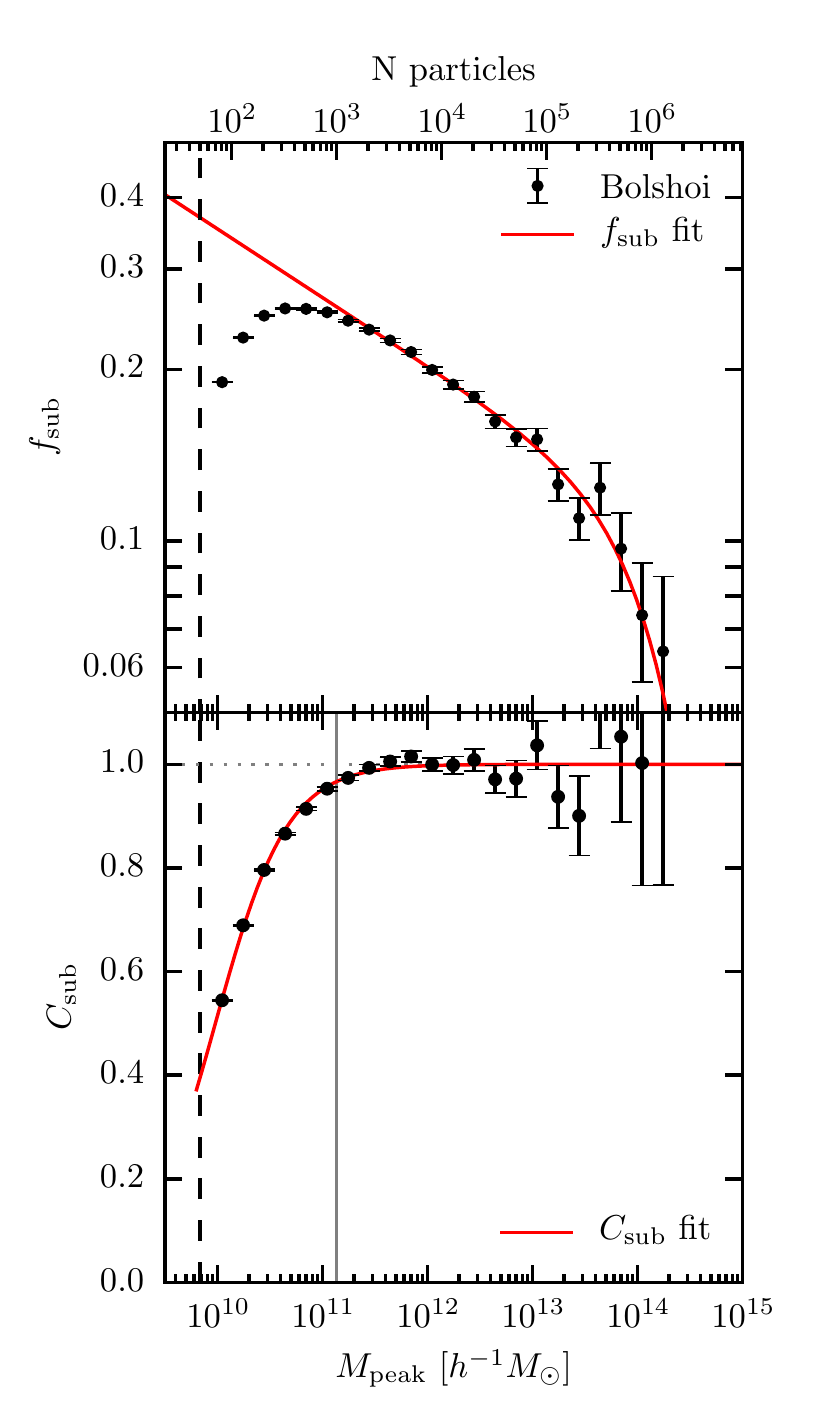}
    \caption{The upper panel shows the fraction of haloes that are sub-haloes, $f_{\rm sub}$, as a function of $M_{\rm peak}$ in the Bolshoi simulation (points with error bars).  This relation is fit with a Schechter function (eq. \ref{eq:fsub_fit}, red line) down to $10^{12} ~ h^{-1}M_{\odot}$ and extrapolated to lower masses.  The lower panel shows an estimate of the sub-halo completeness, $C_{\rm sub}$.  This relation is fit with a function (Eq. \ref{eq:sub_completeness}, red line).   The dashed line marks the 50 particle $M_{\rm peak}$ minimum mass a (sub-)halo must attain to be included in our halo catalogue.  The gray line indicates the 1000 particle mass limit \citet{Guo:2013fm} recommend for convergence in sub-halo abundance.  The upper $x$-axis is the number of particles corresponding to $M_{\rm peak}$ on the lower axis.  The error bars indicate Poisson Errors.}
    \label{fig:f_sub}
\end{figure}

\begin{figure*}
    \includegraphics[width=\textwidth]{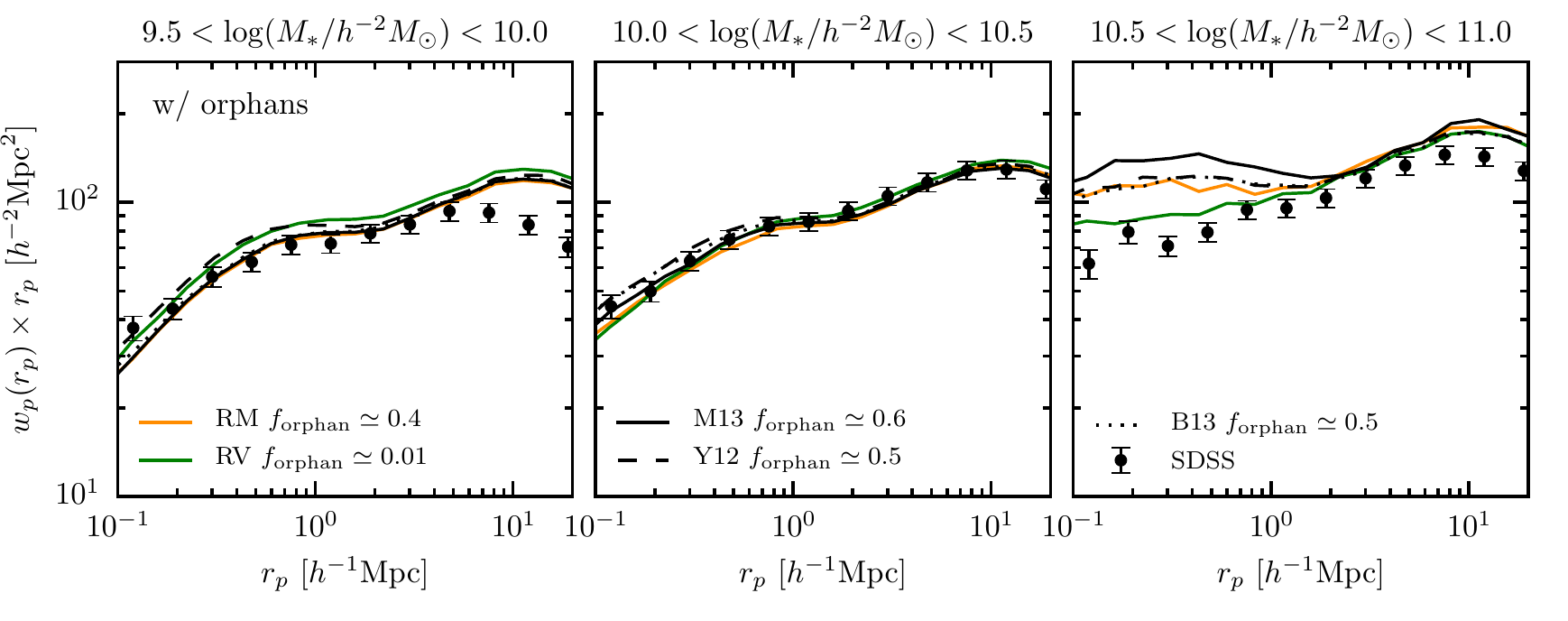}
    \caption{The projected correlation function, $w_p$, is plotted in three stellar mass bins for five SHAM models (lines) where we have introduced orphan galaxies into each model in order to approximately match the clustering in the $\log(M_*)=$[10.0,10.5) mass bin. For comparison, we plot the projected correlation function of galaxies in SDSS as measured by \citet{Yang:2012ew} as points with error bars.  For each model, we list the fraction of satellites that are orphans, $f_{\rm orphan}$.  Compare this to figure \ref{fig:wp_comparison} for the case with no orphans.}
    \label{fig:wp_comparison_w_orphans}
\end{figure*}

\begin{figure}
    \includegraphics[width=\columnwidth]{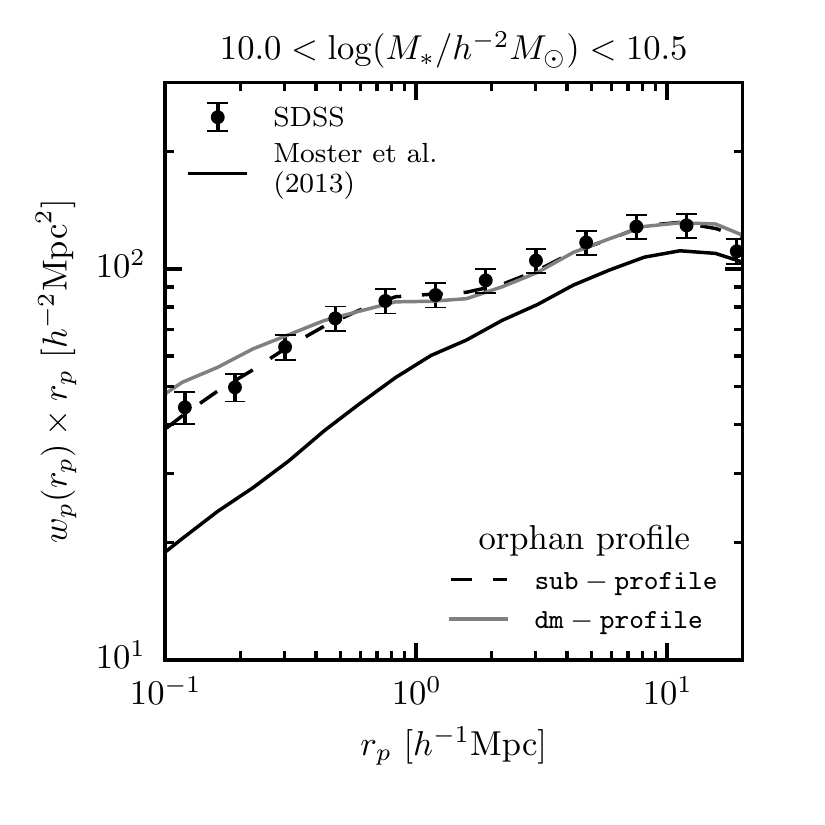}
    \caption{Comparison of the effect of the profile orphan galaxies follow in their host-halo on galaxy clustering.  The solid line shows the clustering prediction for the original M13 model.  The dashed line is M13 including orphans which follow the same profile as extant sub-haloes in the simulation with $C_0=0.6$.  The grey line is the same model but instead orphans are more centrally concentrated, following the profile of dark matter in their host.  See Appendix \ref{sec:clones} for a more detailed discussion.  For comparison, SDSS measurements by \citet{Yang:2012ew} are plotted as points with error bars.}
    \label{fig:orphan_clustering_2}
\end{figure}

One possible solution to the lack of small scale clustering signal in SHAM models is to include a population of ``orphan'' galaxies.  Given the finite mass and force resolution of dark matter simulations, it is reasonable to expect that some sub-haloes are artificially disrupted or otherwise missing from the halo catalogues at $z=0$ \citep[][van den Bosch 2017 in prep.]{Carlberg:1994kl, vanKampen:1995ix, Guo:2013fm}.  Alternatively, sub-halo finders may fail in identifying sub-haloes when the density contrast is low \citep{Wetzel:2010ei, Muldrew:2011gr, Knebe:2011jc, Onions:2012iv, Knebe:2013bp, vandenBosch:2016ky} as is the case in the central regions of host haloes.  If this is in fact occurring, then it is appropriate to include a population of ``orphan'' galaxies, galaxies that have no identifiable sub-halo within a simulation but should rightfully be included if the simulation or sub-halo finder had been more successful.       

After sub-haloes are accreted onto a more massive host-halo, mass is tidally stripped as the sub-halo orbits within the potential of its host, resulting in a ratio between the $z=0$ mass and the mass at accretion that is generally less than unity:
\begin{equation}
f_m = m_{\rm sub}/m_{\rm acc}
\end{equation}
where the mass of a sub-halo, $m_{\rm sub}$, is the instantaneous mass that is bound to the sub-halo, and $m_{\rm acc}$ is the virial mass at the time of accretion.  Eventually, sub-haloes may simply be stripped below the mass resolution of the simulation, $f_m \times m_{\rm acc} \sim \mathcal{O}(10) \times m_{\rm p}$, where $m_{\rm p}$ is the particle mass in the simulation. If it is common for sub-haloes to survive to this point in Nature, this will result in a need for orphans when populating dark matter simulations with galaxies. Alternatively, if sub-haloes are not tracked accurately, mass resolution is not sufficiently high, or the force resolution is not sufficient, some sub-haloes may become disrupted prematurely, meaning fewer sub-haloes will be available to host satellite galaxies when applying a SHAM scheme.

Here we estimate the contribution of this potential missing sub-structure to the abundance of sub-haloes.  In the upper panel of Fig. \ref{fig:f_sub}, we measure the sub-halo fraction as a function of $M_{\rm peak}$ at $z=0$ in 0.2 dex mass bins:
\begin{equation}
\label{eq:fsub}
f_{\rm sub}= \frac{N_{\rm sub}}{N_{\rm sub}+N_{\rm host}}\,.
\end{equation}
\cite{Jiang:2016dw} and \cite{vandenBosch:2016ky} show that the evolved conditional sub-halo mass function is well approximated by a power law with a universal low mass end slope.  Given this, it is expected that $f_{\rm sub}(M_{\rm peak})$ will also be a monotonic power law, increasing towards lower halo masses.

With this in mind, we fit $f_{\rm sub}$ in Bolshoi with a Schechter function of the form:
\begin{equation}
\label{eq:fsub_fit}
f_{\rm sub}(M_{\rm peak}) = f_0 \left( \frac{M_{\rm peak}}{M_0} \right)^{\alpha}e^{-M_{\rm peak}/M_0}\,.
\end{equation}
We find a good fit with $f_0 = 0.105 \pm 0.006$, $\log(M_0/h^{-1}M_{\odot}) = 13.4 \pm 0.1$, and $\alpha = -0.120 \pm 0.005$ as shown as the red line in the upper panel of Fig. \ref{fig:f_sub}.  There is a prominent break in the sub-halo fraction around $10^{11} ~ h^{-1}M_{\odot}$, approximately 1.5 dex above the halo mass identification limit imposed on the simulation, $50 \times m_{p}$.  Therefore, we conservatively use only the measurements above $10^{12}~h^{-1}M_{\odot}$ for our fit to $f_{\rm sub}(M_{\rm peak})$. 

We calculate the sub-halo completeness in the simulation as the ratio of the empirical $f_{\rm sub}$ and the fit using eq. \ref{eq:fsub_fit} as shown in the bottom panel of Fig. \ref{fig:f_sub}.  We then model the completeness, $C_{\rm sub}$, as a function of halo mass as:
\begin{equation}
\label{eq:sub_completeness}
C_{\rm sub}(M_{\rm peak}) = \frac{C_0}{1.0 + \left(\frac{M_0}{M_{\rm peak}}\right)^{\gamma}}
\end{equation}
We find $\log(M_0/h^{-1}M_{\odot})=9.980\pm0.003$ and $\gamma=1.27 \pm 0.02$ provides a good fit, as shown by the red line in the bottom panel in Fig. \ref{fig:f_sub}.  For now, we fix $C_0=1.0$, and we examine the possibility of $C_0<1.0$ at the end of this section.  The implicit assumption here is that when sub-haloes are well resolved, as is the case for massive (sub-haloes) at the time of in-fall, there should be no missing sub-structure.  Our estimation of the completeness is broadly consistent with the 1000 particle threshold found by \citet{Guo:2013fm} using the Millennium simulation suite (shown as the grey line in the bottom panel of Fig. \ref{fig:f_sub}).  Furthermore, we find that $C_{\rm sub}(M_{\rm peak})$ is very nearly constant with redshift (between $z=0-4$).  

The SHAM models considered in this work require sub-haloes in order to populate the simulation with satellites.  To create sub-haloes to host orphan galaxies we ``clone'' extant sub-haloes in the regime where incompleteness results in too few sub-haloes.  Here we briefly describe this process, and we provide a more detailed description in Appendix \ref{sec:clones}.  Where needed, we create a copy of a sub-halo (hereafter `clone') and place it into a new host-halo with approximately the same mass as the donor's host-halo.  We consider two methods for assigning positions and velocities to these new clone sub-haloes that host orphans.  One method conserves the relative position and velocity with respect to the donor's host-halo ({\tt sub-profile}).  The other method assigns the clone the position and velocity of a randomly selected dark matter particle in its new host-halo ({\tt dm-profile}).  We carry over all other relevant properties from donor to clone (e.g. $z_{\rm acc}$). We then add clones into the simulation to make up for incompleteness. Note that we have explicitly assumed that the sub-haloes that host orphans are a fair sampling of surviving sub-haloes in the simulation.  It is possible, even likely, that orphans' sub-haloes are in reality biased in their properties.

We find that when populating the Bolshoi simulation down to $M_* \geq 10^{9.5}~h^{-2}M_{\odot}$ using our model for $C_{\rm sub}$, less than $\sim 1\%$ of satellite galaxies are orphans in each of the SHAM models.  This small orphan percentage suggests that the Bolshoi simulation has sufficient resolution for SHAM studies down to the stellar masses considered in this paper.  Of course, populating these models down to lower masses would result in a larger contribution from orphans.  Nevertheless, we examine the effect of this small orphan population on the galaxy clustering predictions of each SHAM model.  While not shown here, we find that the maximum effect on $w_p$ is of order $\sim 1\%$ on the clustering signal at $0.1~h^{-1}{\rm Mpc}$, regardless of the method used to assign orphan positions in their host.  As expected, the effect of these orphans is even smaller in the more massive stellar mass bins were the $C_{\rm sub}$ correction is smaller.  We conclude that the resolution of the Bolshoi simulation appears to be sufficient for SHAM studies down to the stellar masses considered here and most relevant for galaxy clustering studies using SDSS.    

Finally, we ask ``how many orphans are needed to increase the clustering signal sufficiently in the SHAM models based on $M_{\rm peak}$ to match observations?''  To answer this question, we adjust $C_0$ in eq. (\ref{eq:sub_completeness}), while keeping the other parameters fixed, and fit to the galaxy clustering observations.  Lower values of $C_0$ imply an overall increased population of clone sub-haloes available to host orphans at all masses.  This correction assumes that sub-haloes are being artificially disrupted or merged with the host {\em at all masses}.  For each model, $C_0$ was adjusted in order to best fit the galaxy clustering signal in the intermediate mass bin $(10.0 \leq \log(M_*/h^{-2}M_{\odot}) <10.5)$.  The result of this exercise is shown in Fig. \ref{fig:wp_comparison_w_orphans} using the {\tt sub-profile} method to assign the positions and velocities.  We find that each model requires $C_{0} = [0.6,1.0, 0.4, 0.5, 0.5]$, for the RM, RV, M13, Y12, and B13 models respectively.  While the RV model does not require orphans, the mass-based SHAM models require that on average approximately {\em half of all satellites are orphans}. This number of orphans is within the realm of possibility.  For every surviving sub-halo in Bolshoi, there are $\mathcal{O}(10)$ disrupted sub-haloes (van den Bosch et al. 2017, in prep.).

\citet{Yang:2012ew}, in the original implementation of the Y12 model, used an analytic halo model where sub-halo abundance was a nearly free parameter.  The sub-halo abundances found in that study when clustering observations where fit are consistent with our finding here that $C_0 \sim 0.5$, namely that the sub-halo abundance needs to be approximately a factor of two larger than in the Bolshoi simulation.  A source for these missing sub-haloes is disrupted haloes, which could amount to the factor of 2 if nearly all disruption is artificial (van den Bosch et al. 2017, in prep.).  In fact, with no disruption, galaxy clustering can become stronger than observed on small scales in SHAM models \citep{Watson:2012fu}.  

In Fig. \ref{fig:orphan_clustering_2} we show how assigning orphans' positions within their host affect the clustering predictions for the M13 model in the intermediate mass bin only.  As expected, using {\tt dm-profile} boosts the clustering signal at small scales relative to {\tt sub-profile}, but the effect is small compared to the dependence on $C_{0}$.  This trend is largely similar in the other mass bins considered and for the other SHAM models.  The exact profile followed by orphans is a secondary effect compared to the large abundances required.

We make no attempt to fit for new parameters in the evolving SMHM models using our orphan model and clustering measurements, while the RM and RV models adjust automatically to the increased abundance of sub-haloes.  Adjusting the population of sub-haloes so drastically in the evolving models will have an effect on the parameter inference for the SMHM relation. The stellar mass function in these models changes by $\sim 10\%$.  With this caveat in mind, the most noticeable failure of the mass-based models is the over-prediction of the small scale ($< 1 {\rm Mpc}$) clustering signal in the most massive stellar mass bin ($10.5\leq \log(M_*/h^{-2}M_{\odot})<11.0$).  Each of these models now produce too many massive satellites.  This problem could be reduced by altering the SMHM for massive satellites or reducing the number of massive orphans. It is very easy to imagine that $C_{\rm sub}$ is a function of mass and/or $z_{\rm acc}$ in a way that results in greater completeness of massive sub-haloes.  We leave a detailed study on the self-consistency of including orphan galaxies in SHAM to future work.

\begin{figure}
    \includegraphics[width=\columnwidth]{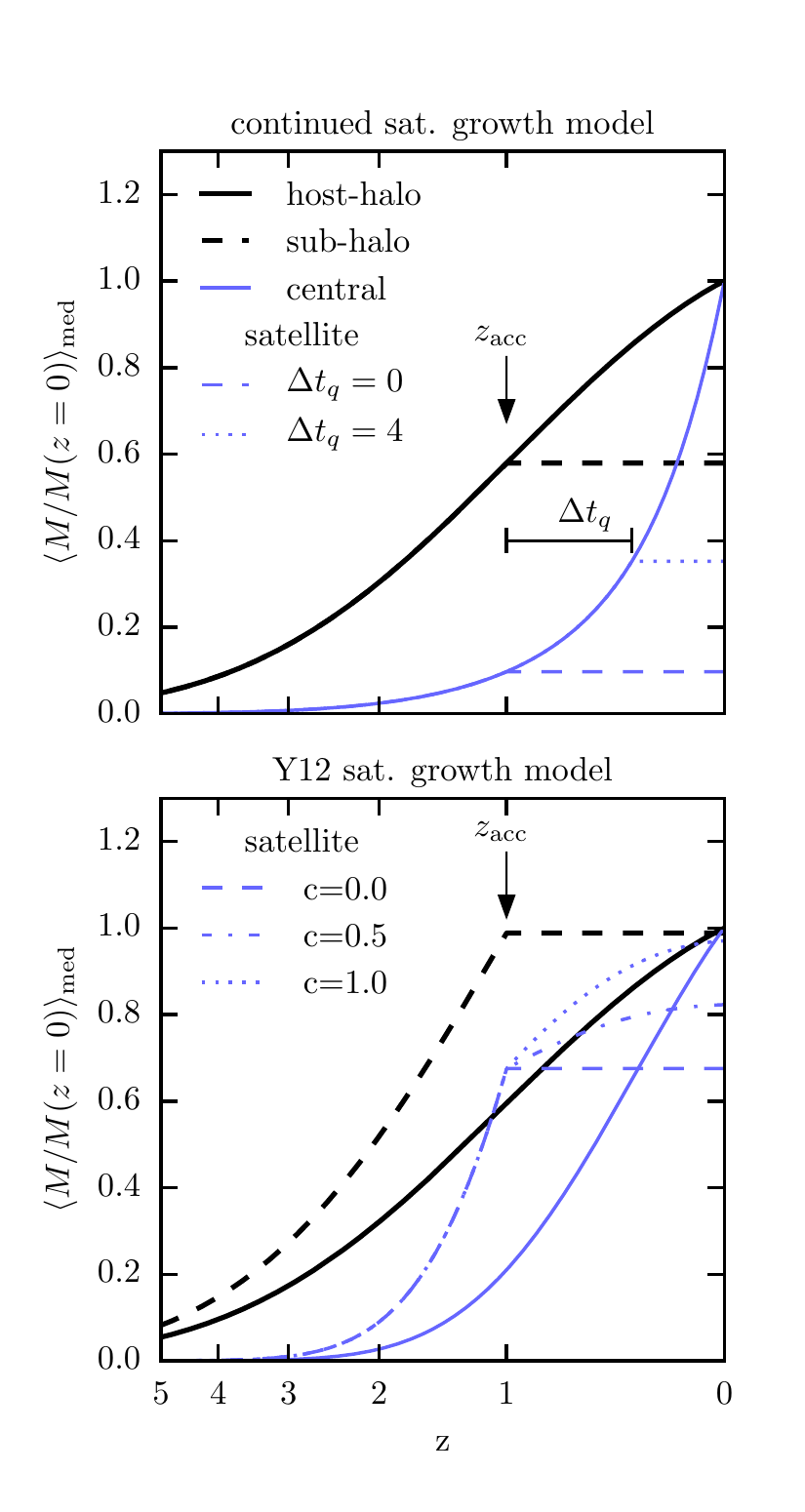}
    \caption{In this figure, we show two models for post-accretion satellite evolution used in this paper.  In both panels, the median mass accretion history (MAH) of an example host-halo (solid black line) and sub-halo accreted at $z=1$ (dashed black line) are shown.  For the sub-halo, we show the peak mass.  Top panel: continued satellite growth model described in \S \ref{sec:sat_growth}.  We show the median stellar mass of a central galaxy which occupies a halo with a $z=1$ mass of $10^{11}~h^{-1}M_{\odot}$ in the M13 model (solid blue line).  The median stellar mass of a satellite for different values of $\Delta t_q$  is indicated by blue dashed and dotted lines for a sub-halo with the same mass at accretion onto an unrelated host-halo.  In each case, the halo (galaxy) growth histories are normalized by the $z=0$ mass of the host-halo (central galaxy).  Bottom panel: the satellite growth model in Y12 described in \S \ref{sec:y12}.  Here, the sub-halo instead has the same mass at accretion as the host-halo at $z=0$.  The growth history of a satellite for different values of the $c$ parameter is shown as various blue lines.}
    \label{fig:sat_growth_model}
\end{figure}

\begin{figure*}
    \includegraphics[width=\textwidth]{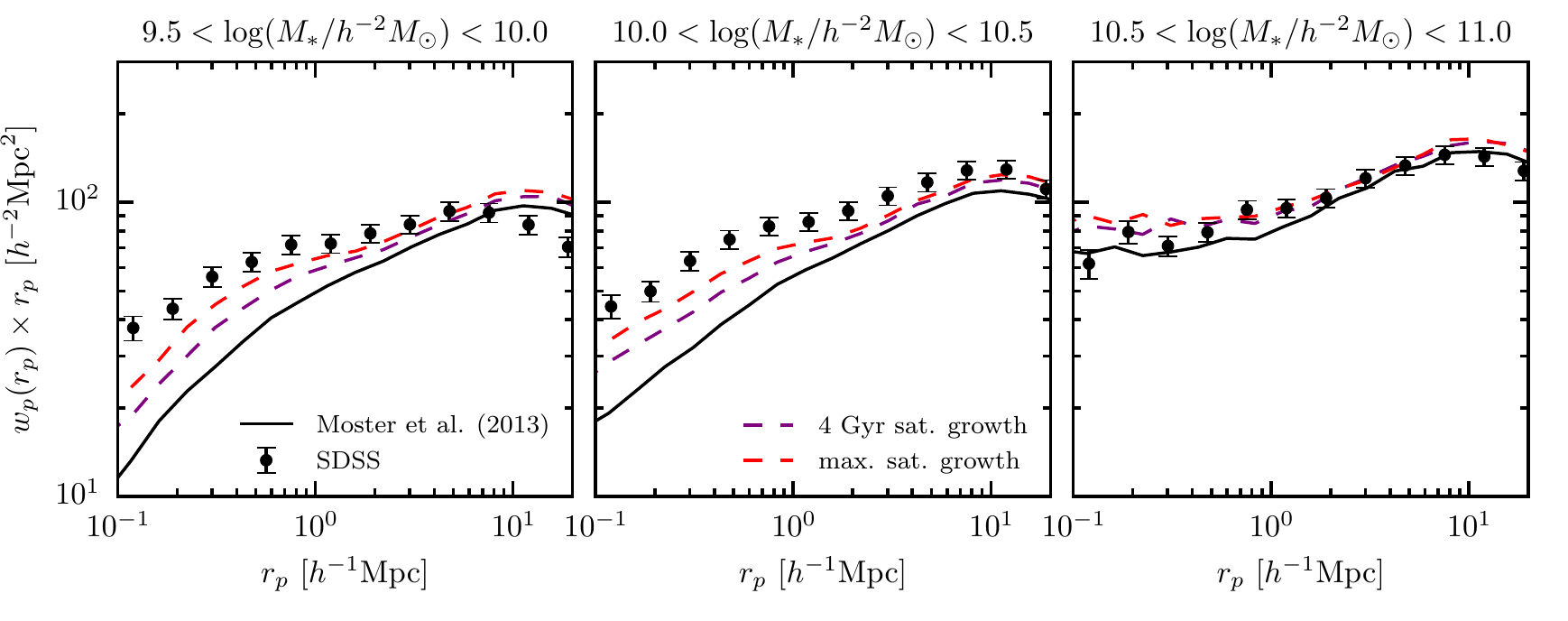}
    \caption{Similar to figure \ref{fig:wp_comparison}.  The original M13 model is shown as a solid line.  The effect of allowing satellites to grow as described in \S \ref{sec:sat_growth} for $4~{\rm Gyr}$ (purple dashed line) and till $z=0$ (red dashed line).}
    \label{fig:wp_effect_comparison_sat_growth}
\end{figure*}

\section{Satellite Growth}
\label{sec:sat_growth}  

While we have shown that a large population of orphan galaxies could alleviate the clustering crisis in mass-based SHAM models, high resolution dark matter simulations do not seem to provide evidence of the requisite missing sub-halo population.  With this in mind, we explore alternative mechanisms to boost the galaxy clustering signal in SHAM models.  In this section, we consider whether continued star-formation in satellites after accretion can significantly boost clustering.  This is motivated by findings suggesting that satellites continue to form stars and grow in stellar mass after accretion for between $\sim 2$ and $4$ Gyr \citep{Wetzel:2013dw}.  If satellites continue to grow after accretion, SHAM methods which use the SMHM relation at $z_{\rm acc}$ to assign stellar mass to satellites (e.g. M13 and B13) will under-estimate $M_*$ in sub-haloes.

It is a common assumption in SHAM models, implicit or explicit, that satellites undergo no significant evolution in stellar mass after $z_{\rm acc}$.  Amongst the models considered in this work,Y12 serves as an exception, explicitly parameterizing post-accretion evolution.  In the bottom panel of Fig. \ref{fig:sat_growth_model}, we provide a graphic in order to explain how growth is parametrized in Y12.  Satellites (broken blue lines) are assigned a stellar mass at z=0 that is between the one achieved at $z_{\rm acc}$ and the one a central galaxy with equivalent $M_{\rm peak}$ at z=0 achieves (solid blue line).  The growth (or mass loss) of satellites is controlled by the ``$c$'' parameter in eq. (\ref{eq:y12sat_model}).  In the case where $c=1$ (dotted blue line), satellites reach the same mass as corresponding central galaxies, and as a result satellites follow the same SMHM relation as centrals at $z=0$, regardless of $z_{\rm acc}$.  In the case where $c=0$ (dashed blue line), the Y12 model is similar to B13 and M13, where there is no evolution in the mass of satellites after $z_{\rm acc}$.  For this work, we set $c = 1$ \citep[consistent with SMF2 FIT-2PCF in table 4 in][]{Yang:2012ew}.  The result of setting $c = 1$ is a model similar to the RM model, where no distinction is made between host-haloes and sub-haloes when abundance matching on $M_{\rm peak}$ at $z=0.0$.  While such a model does produce more massive satellites than the M13 and B13 models, the RM and Y12 models still result in a clustering signal that is too weak on small scales.

In order to further explore the effect of satellite growth, we devise a different model for post accretion evolution similar to that implemented in \cite{Behroozi:2015wx}.  Our primary assumption is that satellites continue to form stars exactly like central galaxies which occupy haloes that had the same mass as the satellite's halo at the time of accretion, $t_{\rm acc}(z_{\rm acc})$, before quenching rapidly after a delay time, $\Delta t_{q}$.  Within the SHAM framework, stellar mass is assigned to (sub-)haloes using a mass proxy, $\mathcal{M}$.  To implement our growth model, we set $\mathcal{M}$ for sub-haloes to the average mass of a host-halo at $t_{\rm acc} + \Delta t_{q}$ which had the same mass as the sub-halo at $t_{\rm acc}$:
\begin{equation}
\mathcal{M} = \langle M_{\rm vir}( t_{\rm acc}+\Delta t_{q}) | M_{\rm vir}( t_{\rm acc}) = M_{\rm acc}\rangle_{\rm med}
\end{equation}
To extrapolate the mass from $t_{\rm acc}$ to $t_{\rm acc}+\Delta t_{q}$ we again use the median MAHs from \citet{Bosch:2014cu}.

The upper panel of Fig. \ref{fig:sat_growth_model} depicts this ``continued satellite growth'' model.  Consider two galaxies, one destined to be a central galaxy at $z=0$, and another that becomes a satellite at $z=1$.  Each galaxy resides in a halo with mass $10^{11}~h^{-1}M_{\odot}$ at $z=1$.  We show the MAH for such a host-halo and sub-halo normalized by the $z=0$ mass of the host-halo as a black solid and dashed line in Fig \ref{fig:sat_growth_model}.  The stellar mass growth history for a central galaxy in such a host-halo in the M13 model is shown as a solid blue line, and the growth history of a satellite galaxy is shown as a long dashed blue line, each  normalized by the stellar mass of the central at $z=0$.  In a model where no evolution occurs post-accretion, the stellar mass of the satellite is set at $z_{\rm acc}$.  An example of the continued growth model is shown as a dotted blue line. Here the satellite continues to grow for $\Delta t_q$ after $z_{\rm acc}$ just as it would have had it remained a central.  In this specific example with $\Delta t_q= 4~{\rm h^{-1}Gyr}$, the satellite's stellar mass increases by 350\% compared to the unmodified M13 model with no post-accretion growth.  In general, the amount a satellite will grow in a fixed $\Delta t_q$ depends on $z_{\rm acc}$, $M_{\rm peak}$, and the evolution of the SMHM relation.  Recently accreted sub-haloes will have less time to grow, and massive sub-haloes will only grow slightly as massive galaxies do not grow rapidly at late times in most evolving models.  

We apply this ``continued satellite growth'' model to the M13 and B13 models.  Initially, we set $\Delta t_{q}=4 ~h^{-1}{\rm Gyr}$, the largest time found by \citet{Wetzel:2013dw}.  As an example, we show the result on the clustering signal for the M13 model in Fig. \ref{fig:wp_effect_comparison_sat_growth} (purple dashed line).  In each model, $\Delta t_{q}=4 ~h^{-1}{\rm Gyr}$ does not result in a sufficient increase in the clustering signal.  To estimate an upper bound on the effect of such a satellite growth model, we allow satellites to grow until $z=0$.  This result is shown in Fig. \ref{fig:wp_effect_comparison_sat_growth} as a red dashed red line.  Even this extreme satellite growth model does not result in strong enough galaxy clustering relative to observations.  The results for the B13 model are very similar. Note that a model in which satellite growth universally continues until $z=0$ is clearly too extreme, as it would predict that satellites have star-formation histories that are indistinguishable from centrals of the same stellar mass, in clear conflict with observations \citep[e.g.][]{Weinmann:2006hu, vandenBosch:2008fv, Wetzel:2012lk, Wetzel:2013dw, Kawinwanichakij:2015tu}.

The failure of a continued growth model for satellites to fit galaxy clustering observations suggest that satellite growth (at least as implemented here) cannot on its own solve the clustering crisis; however, reasonable growth does have a significant effect on the clustering signal at small scales.  This also suggests that the growth model used by Y12 is not sufficient to capture post-accretion growth.  In Y12, a satellite is limited to grow only as massive as a central at $z=0$ with the same peak mass.  In the growth model considered here, satellites may grow more massive than centrals by $z=0$. Any SHAM model that aims to fit galaxy clustering observations will have to confront post-accretion satellite growth; however, it is worth noting that the satellite growth models considered here add significant freedom to otherwise minimally parametrized SHAM models.

\section{Galaxy Assembly Bias}
\label{sec:assem_bias} 

Assembly bias is the phenomenon observed in $\Lambda$CDM simulations of structure formation that the clustering of haloes depends on formation history in addition to mass \citep{Gao:2005ds, Wechsler:2006rg, Zentner:2007bk, Gao:2007yy, li:2008oi, Sunayama:2015ue, Mao:2017ww}.  However, the degree to which the properties of galaxies themselves are influenced by the assembly history of their halo remains an open question, i.e. galaxy assembly bias.  SHAM algorithms that employ measures of $V_{\rm max}$ (like the RV model) already induce assembly bias into galaxies because concentration, and therefore circular velocity, is correlated with formation history at fixed halo mass \citep{Zentner:2014ki,Lehmann:2017fy}.  Conditional abundance matching \citep[CAM,][]{Hearin:2014hh} extends the SHAM framework by allowing for two or more halo properties to influence how galaxy properties are assigned in the SHAM algorithm.  CAM has been used to study the dependence of star-formation rate on halo formation history \citep{Hearin:2013km, Hearin:2014hh, Watson:2015gq, Saito:2015vi, Paranjape:2015uy}; however, SHAM models that assign stellar mass and star-formation in a self consistent manner are still in development.

\begin{figure}
    \includegraphics[width=\columnwidth]{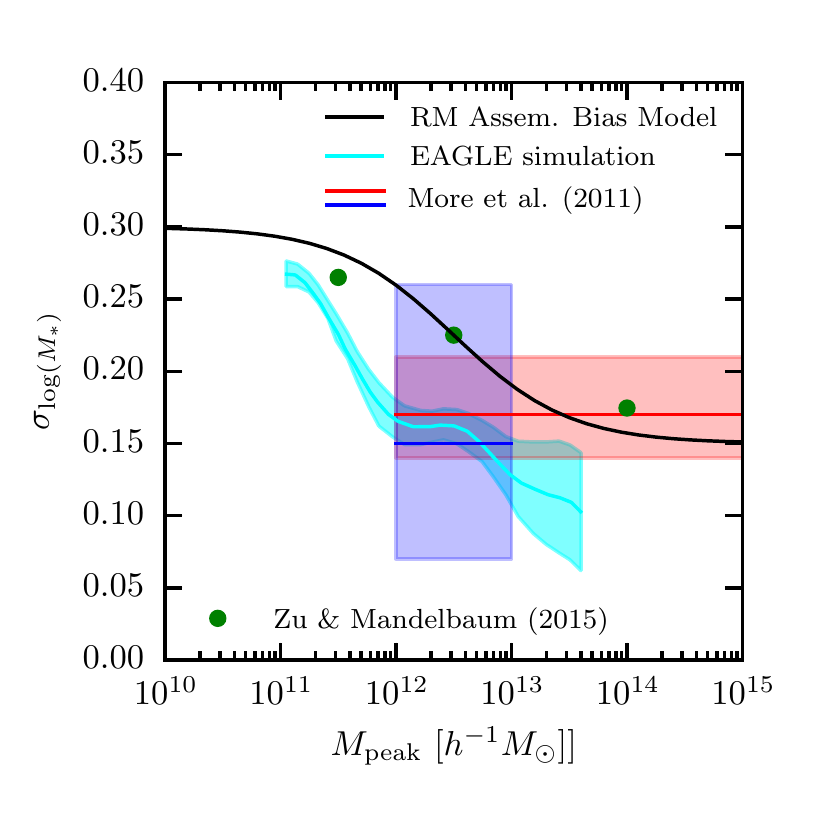}
    \caption{Here we plot the log-normal scatter in stellar mass as a function of peak halo mass for our model as a solid black line.  For comparison, we show various measurements and theoretical prediction for the scatter.  The cyan line and shaded region shows the relation for central galaxies in the EAGLE simulation with the 1$\sigma$ errors \citep{Matthee:2016vm}.  The red and blue lines with shaded regions show the fixed scatter determined for red and blue central galaxies and the associated error measured from satellite kinematics \citep{More:2011il}.  The green points show the scatter at three masses determined from an HOD analysis with weak lensing measurements \citep{Zu:2015vh}.}
    \label{fig:scatter_model}
\end{figure}

\begin{figure}
    \includegraphics[width=\columnwidth]{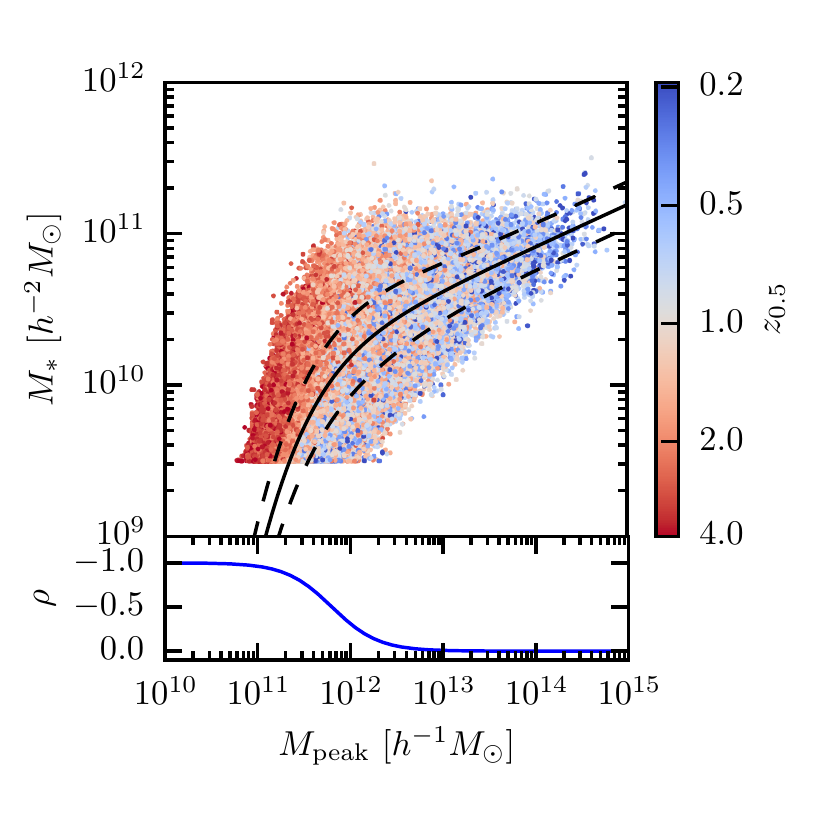}
    \caption{In the upper panel, we plot the SMHM relation for individual galaxies as points, colour-coded by the formation redshift, $z_{0.5}$.  The mean relation is plotted as a solid line with the $\pm 1\sigma$ scatter shown as dashed lines.  In the bottom panel, we show the strength of assembly bias, parametrized by the $\rho$ parameter.}
    \label{fig:assem_bias_smhm}
\end{figure} 

\begin{figure*}
    \includegraphics[width=\textwidth]{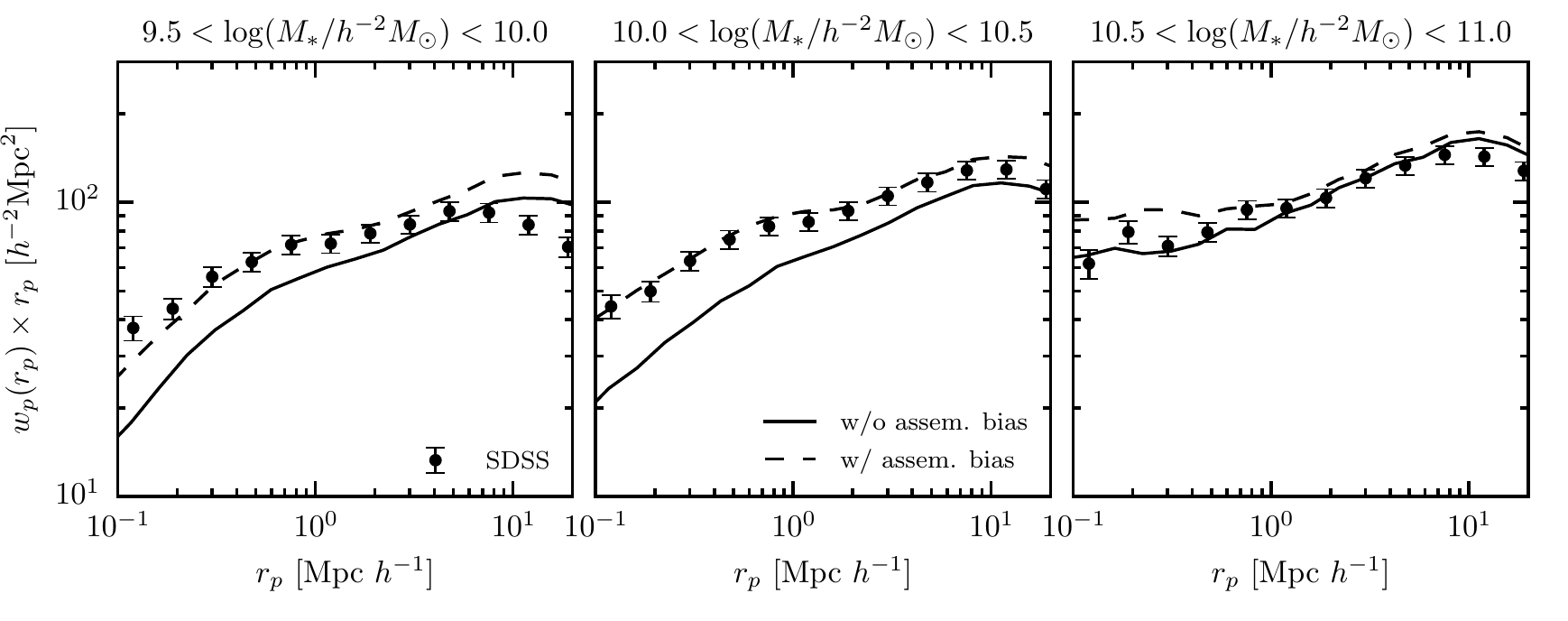}
    \caption{Similar to figure \ref{fig:wp_comparison}.  The original RM model is shown as a solid line.  The combined effect of including assembly bias with scatter in the SMHM relation in the RM model is shown as a dashed line.}
    \label{fig:wp_comparison_assem_bias}
\end{figure*}

In this section we examine whether introducing assembly bias into the galaxy-halo mapping can increase the satellite fraction and therefore the galaxy clustering signal in mass-based SHAM models.  Specifically, we consider a model where $M_{*}$ is correlated with the formation history of the (sub-)halo it occupies such that early forming haloes host more massive galaxies than late forming haloes at {\em fixed} peak (sub-)halo mass.  While many measures of halo formation history have been employed in the literature, for this work we use the redshift at which a halo first achieves a fraction, $f$, of its peak halo mass, $z_{f}$ (see Appendix \ref{sec:halo_properties} for details on how $z_{f}$ is calculated).

\subsection{Rank Order SHAM Assembly Bias}

We begin by modifying the RM model to include galaxy assembly bias.  In order for assembly bias to have any effect on $M_*$, there must be a significant amount of scatter in the SMHM relation, $\sigma_{\log(M_*)} >0$.  Scatter provides a dynamic range in $M_*$ at fixed $M_{\rm peak}$ over which $M_*$ can be correlated with formation history.  In Fig. \ref{fig:scatter_model}, we shown some constraints on $\sigma_{\log(M_*)}$ from the literature.   Typical values found for $\sigma_{\log(M_*)}$ in models that do not include assembly bias are between $0.1-0.2$ dex \citep{More:2011il, Reddick:2013gi, Zu:2015vh, Zentner:2016tz, Tinker:2016vu, Lehmann:2017fy}; however, these values are most strongly constrained at high masses, $M_{\rm vir}>10^{12}~h^{-1}M_{\odot}$.  We add scatter to the SMHM in the RM model using the method from \citet{Behroozi:2010ja}, parameterizing the level of scatter as a function of halo mass, $\sigma_{\log(M_*)}(M_{\rm peak})$.   

To induce a correlation between $M_{*}$ and $z_{f}$, we apply the CAM method by binning (sub-)haloes in small 0.1 dex $M_{\rm peak}$ bins. We then rank order (sub-)haloes by $z_{f}$ and galaxies by $M_{*}$, re-assigning the most massive galaxies to the earliest forming (sub-)haloes in the bin.  We use ${f}=0.5$, the redshift where a (sub-)halo reaches half its peak mass.  Furthermore, we parametrize the strength of this correlation by the Spearman rank order correlation coefficient, $\rho(M_{\rm vir})$.  We are able to weaken the correlation by degrading the rank ordering as described in appendix \ref{appendix:ranks} in order to reduce the effect of assembly bias on $M_*$.  In this way, we can set the correlation strength as a function of stellar mass, $\rho(M_*)$.

Initially we try a model with fixed scatter, $\sigma_{\log(M_*)} = 0.18$ dex (similar to the other mass-based models), and a constant maximum correlation ($\rho=1$) between $M_{*}$ and $z_f$. This model produces a poor fit to the observed galaxy clustering signal.  The satellite fraction of massive galaxies in such a model is unrealistically large.  As a result, the clustering signal increases dramatically at high stellar masses, while at the same time, there is relatively little effect on the clustering signal at the lower stellar masses where the clustering is under-predicted.  This can be understood as a result of the SMHM relation becoming steep at the low mass end, and $\sim 0.18$ dex scatter in the SMHM relation is not sufficient to increase the satellite fraction significantly.

With this in mind, we modify our model to fit the observed galaxy clustering signal by changing the parametrization of $\sigma_{\log(M_*)}$ and $\rho(M_{\rm vir})$.  First, we require the scatter to increase at lower halo masses ($<10^{12}h^{-1}M_{\odot}$) and decrease at high halo masses ($>10^{12}h^{-1}M_{\odot}$) where there are observational constraints.  While there are few constraints on the scatter in low mass haloes, there are some indications that it may in increase as mass decreases (see Fig. \ref{fig:scatter_model}). However, it should be noted that most previous studies that constrain scatter do so assuming no galaxy assembly bias. Including assembly bias may allow for significantly larger amounts of scatter \citep{Zentner:2014ki, Zentner:2016tz}.  

We model the dependence of $\sigma_{\log(M_*)}$ on $M_{\rm peak}$ in the RM model as:
\begin{equation}
\label{eq:sigma_mstar}
\sigma_{\log(M*)}(M_{\rm peak}) = f(M_{\rm peak})
\end{equation}
where $f(x)$ is a sigmoid function of the form,
\begin{equation}
f(x) = \frac{y_1 - y_0}{1+ e^{k(x-x_0)}} + y_0
\end{equation}
This function asymptotes to $y_0$ for $x \gg x_0$ and $y_1$ for $x \ll x_0$.  Second, we require the effect of assembly bias to be minimal at high masses, where there is no need to increase the clustering signal, and stronger at low masses.  Similar to the model for scatter, we parametrize the strength of assembly bias, $\rho$, as a function of halo mass using the same functional form:
\begin{equation}
\label{eq:rho_mpeak}
\rho(M_{\rm peak}) = f(M_{\rm peak})
\end{equation}
By experimentation, we find parameters for the scatter model (Eq. \ref{eq:sigma_mstar}): $\log(x_{0})=12.0$, $y_1=0.3$, $y_0=0.15$, and $k=2.0$, and the assembly bias strength model (Eq. \ref{eq:rho_mpeak}): $\log(x_{0})=11.8$, $y_1=-1.0$, $y_0=0.0$, and $k=4.0$, provide a good fit to the SDSS galaxy clustering observations.  The model for scatter is broadly within the range of values found in other studies (shown as a black line in Fig. \ref{fig:scatter_model}).  The resulting SMHM relation in the RM model is shown in the top panel of Fig. \ref{fig:assem_bias_smhm}, with the strength of galaxy assembly bias shown in the bottom panel.  

The clustering signal in the RM model with assembly bias is shown in Fig. \ref{fig:wp_comparison_assem_bias} along with the original RM model with no assembly bias effect.  The model with assembly bias is much more consistent with galaxy clustering observations.  In addition, the satellite fraction is very similar to the RV model.  The success of this scheme to fit clustering observations (and the RV model) suggests that assembly bias can increase the satellite fraction and therefore the clustering signal in SHAM models. This is not surprising given that the RV model, which fits the clustering, is essentially an example of such an assembly bias model.  However, neither of these models provide a {\em solution} to the crisis in this paper.  Neither self consistently model the evolution of the stellar mass function.  In addition, the formation redshift of sub-haloes is compared to host-haloes at $z=0$ regardless of $z_{\rm acc}$.  A more physical approach would be to compare sub-haloes to other host-halos at $z_{\rm acc}$ when assigning stellar mass at accretion. In the next section, we attempt to self-consistently add assembly bias to the evolving SHAM models.    

\subsection{Evolving SHAM Assembly Bias}

Given the success of our modifications to the RM model to introduce assembly bias to $M_*$, we now consider modifications to the evolving mass-based modes.  As discussed, each of the evolving models makes the assumption that satellite galaxies at the time of accretion have the same mass as central galaxies in haloes with equal $M_{\rm peak}$.  In these models, correlating $M_*$ with $z_{\rm f}$ will increase the mass of satellite galaxies if haloes which become sub-haloes form significantly earlier than haloes which remain host-haloes.  This differs from the model for post-accretion satellite growth discussed in \S \ref{sec:sat_growth}.  In this model, galaxies which become satellites grow more quickly than galaxies which remain centrals.  Therefore, satellites are ``over-massive'' compared to centrals at $z_{\rm acc}$. 

For the evolving models we adopt the same scatter model used for the RM model as described above.  Instead of applying the CAM method at only $z=0$, we apply it at each snapshot, $z_{\rm snap}$, of the simulation, storing the $M_*$ assigned to satellites that were just accreted, i.e. $z_{\rm acc} = z_{\rm snap}$.  In this way, the formation time of satellites is compared to host-haloes at $z_{\rm acc}$.  After $z_{\rm acc}$, satellites are assumed to not grow further.  This is accomplished in a computationally efficient manner as follows.  For each (sub-)halo, we compare its formation time, $z_{\rm f}$, to the full $p(z_{\rm f} | M_{\rm peak}, z_{\rm acc})$ of all host-haloes of equivalent mass at the redshift of accretion (for host haloes we set $z_{\rm acc}=0$) to find its associated percentile location in the distribution, $p_{\rm f}$.  We then assign stellar mass to (sub-)haloes by modifying Eq. \ref{eq:mstar_assign} such that the scatter is now correlated with $z_{\rm f}$:
\begin{align}
\log[M_*(M_{\rm peak}, a_{\rm acc})] =& \log[\langle M_* | M_{\rm peak}\rangle(a_{\rm acc})] \\ \nonumber
&+ \mathcal{F}^{-1}(0,\sigma_{\log(M_*)}, 1-p_{\rm f})
\end{align}
where $\mathcal{F}^{-1}(0,\sigma_{\log(M_*)}, 1-p_{\rm f})$ is the quantile function of a normal distribution with mean 0 and log-scatter $\sigma_{\log(M_*)}$.  In this way, galaxies with earlier formation times are assigned larger stellar masses. 

We find that no formation time parameter between $z_{0.1}$ and $z_{0.9}$ is sufficient to increase $M_*$ of satellites enough to affect the clustering signal in each model substantially enough to fit the observed clustering signal. The earliest formation time we try in this model is $z_{0.1}$, which results in the weakest effect.  The most recent formation redshift we try is $z_{0.9}$, which results in the strongest effect. However, even using $z_{0.9}$ only increases $f_{\rm sat}$ in the B13 and M13 models by $\sim 2-3\%$, resulting in a minimal effect on galaxy clustering.  The primary reason for this is that haloes which become sub-haloes do not have sufficiently earlier formation redshifts compared to haloes which remain host-haloes.  However, since we only examine $z_f$ as the secondary halo property, it remains possible that there exists a halo property which is better correlated with becoming a sub-halo. 

\section{Discussion \& Summary}
\label{sec:discussion}

We have shown that there is no published sub-halo abundance matching (SHAM) model that simultaneously:
\begin{enumerate}
\item fits the clustering of galaxies at $z=0$, $w_p(r_p)$,
\item reproduces the evolution of the stellar mass function, $\phi(M_*,z)$,
\item and uses only identified, extant, sub-haloes in high resolution dark matter simulations.
\end{enumerate}
Models that fit observations of galaxy clustering are not obviously compatible with the observed evolution of the stellar mass function and expectations for the build up of stellar mass in haloes. Conversely, SHAM models which self-consistently fit the stellar mass function as it evolves significantly under-predict galaxy clustering signals at small scales ($\leq 1~h^{-1}{\rm Mpc}$). This tension exposes a clustering ``crisis'' for mass-based SHAM.   

Of the five different models examined in this work, only SHAM based on rank ordering (sub-)haloes by peak maximum circular velocity, $V_{\rm peak}$, (RV  model) results in a robust galaxy clustering signal consistent with observations.  This finding is in-line with previous work that finds $V_{\rm peak}$ is the best quantity to use to reproduce galaxy clustering observations \citep{Reddick:2013gi, Lehmann:2017fy}.  However, if one considers $V_{\rm peak}$ as the best physical quantity that predicts stellar mass, an implicit assumption in $V_{\rm peak}$-based SHAM is that the stellar mass-$V_{\rm peak}$ (SMVP) relation does not evolve.  Because haloes grow their potential wells early \citep[e.g.][]{Bosch:2014cu}, a non-evolving SMVP relation results in galaxies forming too early.  It is not clear how to square this result with analysis of the EAGLE simulation \citep{Schaye:2015gk} which suggests $V_{\rm max}$-related quantities are truly more fundamental for determining stellar mass \citep{ChavesMontero:2015dc, Matthee:2016vm}.

SHAM models based on peak halo mass, $M_{\rm peak}$, (RM, M13, B13, and Y12 models) do not produce strong enough clustering signals with respect to observations, especially on small scales.  Evolving models like M13, B13, Y12, and more recently by \citet{RodriguezPuebla:2017uo}, have been used to learn about the star-formation history of galaxies, quenching physics, and the contribution of merging to the build up of galaxies and stellar haloes with significant success.  Given the wide-ranging utility of these models, we have examined three modifications to mass-based SHAM models that alleviate the clustering crisis to varying degrees: the addition of orphan galaxies, post-accretion stellar mass growth of satellites, and galaxy assembly bias.  Each of these ``solutions'' addresses the clustering crisis by increasing the number of satellite galaxies.

This tension between fitting clustering observations and abundance of satellites is what drives the demand for increasing the number of satellites in many other studies.  For example, in the original semi-analytic implementation of the Y12 model, \cite{Yang:2012ew} find a need for more massive satellites {\em and} longer sub-halo survival times than traditional SHAM implementations.  It is only when we apply the Y12 model to dark matter simulations directly, that it becomes clear that there are not enough extant sub-haloes to fit clustering observations.  \citet{Lim:2016ul} find a similar result when fitting the conditional stellar mass function (CSMF), i.e. satellite abundances.  Only models which allow for a significant orphan population can provide a good fit the the faint end of the CSMF, another indication that the true culprit in the clustering crisis is a lack of satellite galaxies.  Furthermore, this problem is not unique to SHAM models.  \cite{Pujol:2017ua}, in a comparison of many galaxy formation models, find that only models with orphan galaxies are able to fit clustering observations on small scales. 

While the need for orphan galaxies is well established, the motivation for the missing sub-haloes associated with orphan galaxies in high resolution simulations is lacking.  While the mass resolution of simulations places an absolute limit on their ability to resolve highly stripped sub-haloes (no structures can exist below the particle mass, $m_p$), we find no empirical evidence of sufficiently massive missing sub-haloes.  By examining the power-law behaviour of the sub-halo fraction, we find that the resolution of the Bolshoi simulation, $m_{\rm p} = 1.35 \times 10^8 \msunh$, appears to be sufficient to model SDSS-like galaxy samples ($M_* \geq 10^{9.5}~h^{-2}M_{\odot}$), where the majority of satellite galaxies live in haloes with peak masses $\gtrsim 1000 \times m_p$.  This finding is consistent with an independent analysis based on the convergence of the galaxy clustering signal in SHAM models \citep{Guo:2013fm}.  Despite the lack of evidence for large numbers of missing sub-haloes, we find that mass-based SHAM models require that approximately {\em half} of all satellite galaxies are orphans in order to fit galaxy clustering observations.  This large fraction of orphan galaxies is similar to the number required by \cite{Yang:2012ew}.      

Regardless, the appeal of SHAM is based on its ability to leverage the statistical power of large, cosmological, dark matter only (DMO) simulations.  If DMO simulations are not able to resolve substructure abundance to within a factor of $\sim 2$, the utility of SHAM becomes questionable.  Furthermore, DMO simulations may not be reliable probes of substructure if the presence of baryons and various astrophysical processes associated with galaxy evolution significantly modify the abundance, distribution, and structure of sub-haloes.  For example, the inclusion of baryons in cosmological simulations may more tightly bind sub-haloes, therefore increasing the survival time and abundance \citep{Fiacconi:2016ba} relative to DMO simulations.  However, the net effect of baryonic physics on sub-haloes is not well understood.  \citet{Despali:2016uv} find that the abundance of sub-haloes with peak mass $\sim 10^{10}~h^{-1}M_{\odot}$ is increased in the EAGLE simulation, while it is decreased in the Illustris simulation \citep{Vogelsberger:2014gw}.  An enhanced destruction of dwarf galaxy mass sub-haloes ($10^{5} -10^{10} ~h^{-1}M_{\odot}$) has been found in many zoom-in simulations \citep{Read:2006da, Read:2006hw, Brooks:2014jv, Wetzel:2016iy}.  \citet{GarrisonKimmel:2017tu} find that the tidal field of central galaxies' disks results in a depletion in the abundance of sub-haloes by a factor of $\sim 2$ in the central regions compared to DMO simulations.  If the inclusion of the baryonic physics of galaxy formation and evolution generically decreases the abundance of sub-haloes, this only serves to increase the small scale galaxy clustering problem in SHAM.  

Given the uncertain contribution of orphan galaxies, in this paper we have also examined two other physically motivated methods to enhance the satellite contribution in mass-based SHAM models.  First, we examine the effect of allowing satellite galaxies to grow in mass after accretion for some time before quenching.  Within this framework, the process(es) which quenches satellites is delayed, while in the interim satellites continue to form stars similarly to central galaxies \citep{Wetzel:2013dw}.  This idea is at odds with the assumption in many SHAM models that the stellar mass of satellites is set at $z_{\rm acc}$ and serves as a sort of fossil record of the SMHM relation at that redshift.  Continued growth after accretion generally increases the number of satellites above a given stellar mass threshold.  We find that reasonable delay times before quenching result in modest increases to the satellite fraction and, as a result, the clustering signal on small scales.  Again, Y12 find evidence for significant post-accretion evolution of stellar mass, such that satellites acquire a stellar mass that is close to that of central galaxies.  Our model for growth allows for even larger masses, but remains insufficient.  \citet{Behroozi:2015wx} apply a similar model for post-accretion growth and find consistent results when examining close galaxy pairs, but do not comment on galaxy clustering.  Recent work by \citep{Moster:2017vz} suggests post-accretion satellite growth is very important in order to reproduce small scale clustering, although orphans are still needed.  Regardless, our results suggest post-accretion evolution of satellites is an important phenomena to model in order to reproduce the small scale clustering of galaxies, but this effect on its own is not sufficient to solve the clustering crisis in this paper.  Interestingly, in the EAGLE simulation \citet{ChavesMontero:2015dc} find that significant growth occurs after accretion for satellites, and this contributes significantly to increasing the clustering signal on small scales, in agreement with empirical results.         

Finally, we show that galaxy assembly bias can increase clustering in mass-based SHAM models.  The increased clustering signal in $V_{\rm peak}$-based SHAM is a result of assembly bias \citep{Zentner:2014ki,Lehmann:2017fy}.  Using the CAM technique, we show that $M_{\rm peak}$-based SHAM can produce similar results if it assumed that stellar mass is correlated with halo formation time at fixed $M_{\rm peak}$.  We find that such a model must contain two features.  First, the strength of the galaxy assembly bias must decrease in high mass haloes.  Second, the scatter in the SMHM relation must increase towards lower masses.  Both of these features are consistent with features found in the EAGLE simulation \citep{Matthee:2016vm}.  $M_{\rm peak}$-based SHAM with galaxy assembly bias explicitly added appears very similar to $V_{\rm peak}$-based SHAM; however, neither model offers a consistent picture of how (sub-)haloes build stellar mass.  As a result, this is not a complete solution to the crisis in this paper.  

The ability of rank order SHAM models to fit galaxy clustering observations when galaxies are affected by assembly bias motivates the construction of self-consistent evolving SHAM models with a similar assembly bias effect.  In such a model, we assume that galaxies which become satellites have grown more massive by the time they are accreted, compared to galaxies in haloes of similar mass that remain centrals. After accretion, it is assumed that satellites consolidate their stellar mass.  By correlating stellar mass and formation time {\em at the time of accretion} for satellites at every redshift output in the simulation, we compare the formation time of those haloes being accreted during the redshift interval in question to that of all other (host) haloes.  Unfortunately, the difference between the formation time of sub-haloes and host-haloes when measured in this way is relatively small. As a result, the effect of assembly bias is also very small, and the evolving models with assembly bias continue to predict clustering signals that are too weak. That being said, we have only explored one type of assembly bias, and it remains to be seen whether other halo parameters exist that are more correlated with the chance of a halo becoming a sub-halo.

Regardless of the method, matching the small scale clustering with SHAM requires that satellite galaxies at $z=0$ are more massive than central galaxies in haloes of equal peak mass, unless one allows for
orphans. Matching the detailed or even aggregate stellar mass growth history of both central and satellite galaxies may be beyond simple one (or two parameter) SHAM (CAM) models.  We speculate that a combination of both continued stellar mass growth after accretion and galaxy assembly bias are necessary to resolve this crisis.

\section*{Acknowledgements}

DC would like to thank Andrew Hearin for substantive discussions and collaborative work on {\tt Halotools} which both facilitated and greatly improved this paper, and Nir Mandelker, Allison Merritt, and Jeremy Bradford for helpful comments throughout.  This work makes significant use of the {\tt Python} packages and libraries: {\tt NumPy} \citep{VanDerWalt:2011dp}, {\tt IPython} \citep{Perez:2007jl}, {\tt Matplotlib} \citep{Hunter:2007jl}, and {\tt SciPy} \citep{Jones:2001uv}. FvdB is supported by the Klaus Tschira Foundation and by the US National Science Foundation through grant AST 1516962. NP is supported in part by DOE DE-SC0008080. YYM is supported by the Samuel P.\ Langley Postdoctoral Fellowship at the Pittsburgh Particle physics, Astrophysics, and Cosmology Center (Pitt PACC) a the University of Pittsburgh. ASV and ARZ are supported by grants AST 1516266 and AST 1517563 from the U.S.\ National Science Foundation (NSF) as well as by Pitt PACC. JUL was supported by a KITP graduate fellowship and by the US National Science Foundation through grant AST 1516962.

\bibliographystyle{mnras}
\bibliography{bib} 

\appendix

\section{Halo Properties}
\label{sec:halo_properties}

In this section we describe how we calculate properties for (sub-)haloes which depend on their growth history.  We use the merger trees constructed using the {\tt Consistent} trees algorithm \citep{Behroozi:2013dz} built on the {\tt ROCKSTAR} halo catalogues from the Bolshoi simulation.  We distinguish between host-haloes and sub-haloes using the {\tt upid} tag for each halo.  If $ {\tt upid} \equiv -1$, a halo is considered a host; otherwise, if ${\tt upid} > 0$, we consider it a sub-halo.  For this work, we do not distinguish between higher order sub-haloes (i.e. sub-sub-haloes).

\begin{figure*}
    \includegraphics[width=\textwidth]{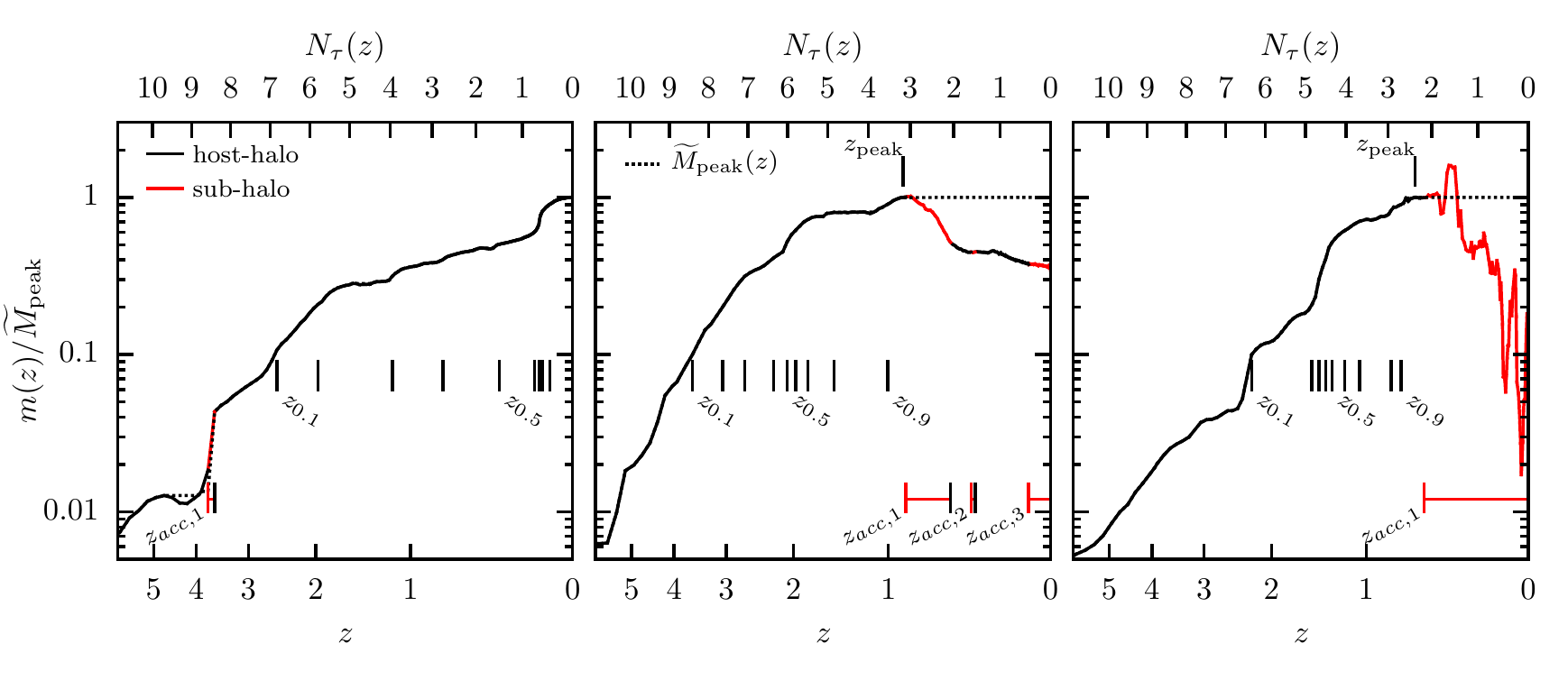}
    \caption{As an example, we show the mass growth histories for three haloes, all with a peak mass of $\sim 10^{12}~h^{-1}M_{\odot}$.  When a halo is identified a host-halo (${\tt upid} \equiv -1$) $m_{\rm vir}(z)/\widetilde{M}_{\rm peak}$ is shown as a solid black line.  When a halo is identified as a sub-halo (${\tt upid} > 0$) $m_{\rm vir}(z)/\widetilde{M}_{\rm peak}$ is shown as a solid red line.  A series of formation times, $z_{f} = [0.1,0.2,...,0.9]$ are marked as vertical black dashes below the growth histories.  Similarly, $z_{\rm peak}$ is marked above the growth histories.  At the bottom of each panel, any accretion redshifts are shown as a red vertical dash connected to the associated ejection redshift marked with a black vertical dash (or through z=0 if it remains a sub-halo).  The running peak mass, $\widetilde{M}_{\rm peak}(z)$, is plotted as a dotted black line.  The upper x-axis in each panel is the number of dynamical times until z=0 from eq. \ref{eq:t_dyn}.}
    \label{fig:growth_hist}
\end{figure*}

\subsection{Peak Halo Mass}

We use the peak halo mass obtained by each (sub-)halo to assign stellar mass in each of the models discussed in this paper.  We calculate the peak mass, $\widetilde{M}_{\rm peak}$, a (sub-)halo at z=0 obtained throughout its history while not identified as a sub-halo as:
\begin{equation}
\widetilde{M}_{\rm peak} = {\rm MAX}[m^{\prime}_{\rm vir}(z)]
\end{equation}
where,
\begin{equation}
m^{\prime}_{\rm vir}(z)= 
\begin{cases}
    m_{\rm vir}(z)   & \quad \text{if host-halo at z}\\
    0.0                   & \quad \text{if sub-halo at z} \\
  \end{cases}
\end{equation}
We then define $z_{\rm peak}$ as the redshift where $m^{\prime}_{\rm vir}(z_{\rm peak})\equiv \widetilde{M}_{\rm peak}$.  This differs from the typical definition of peak halo mass which does not require the peak mass to be obtained while a halo is identified as a host-halo:
\begin{equation}
M_{\rm peak} = {\rm MAX}[m_{\rm vir}(z)]
\end{equation}
The former definition, $\widetilde{M}_{\rm peak}$, disregards any mass growth which occurs while a halo is identified as a sub-halo. We prefer $\widetilde{M}_{\rm peak}$ as a physical parameter because most growth that occurs while a halo is identified as a sub-halo is most often a numerical artifact.  However, we do note that we ignore the rare case of subhalo-subhalo mergers.  In Fig. \ref{fig:growth_hist} we show the growth histories for three haloes in the Bolshoi simulation.  In each panel, we also show the running $\widetilde{M}_{\rm peak}(z)$ and mark the redshift where $\widetilde{M}_{\rm peak}$ is reached.  In the right hand panel we show a case where $\widetilde{M}_{\rm peak} \neq {M}_{\rm peak}$ for a halo which briefly `grows' in mass after accretion.  We find this is the case for $\sim 10\%$ of sub-haloes.

\subsection{Halo Accretion Time}
\label{sec:halo_acc_time}

The purpose of this section is to define a ``primary'' accretion redshift, $z_{\rm acc, prim}$, which is most important for galaxy evolution.  Each of the evolving models in this work require a single accretion redshift for all sub-haloes, where it is assumed that the stellar mass of satellites is set at the time of accretion, or where a special post-accretion growth regime begins.  However, a halo may undergo many accretion events throughout its history.  This makes the identification of a single, most important, accretion redshift non-trivial.

With this in mind, we define the accretion redshift of a halo as the redshift at which it is first identified as a sub-halo after having been first identified as a host-halo\footnote{For the rare case of `immaculate' sub-haloes, sub-haloes with no progenitor \citep{vandenBosch:2017fj}, we use the first redshift for which the halo is identified.}.  Furthermore, because a sub-halo's orbit may take it beyond the virial radius of its host (e.g. backsplash haloes), it is possible to identify multiple accretion redshifts for many haloes.  Given this, we define $z_{\rm acc,n}$ as the redshift a halo is identified as a sub-halo for the $n^{\rm th}$ time.  We also tabulate `ejection' redshifts, $z_{\rm eject, n}$, the redshift a halo is identified as being a host-halo after previously having been identified as a sub-halo for the $n^{\rm th}$ time.  As an example, in the middle panel of Fig. \ref{fig:growth_hist}, we show the growth history of a halo which underwent three accretion events and two ejections since $z \sim 6$.

We explore four definitions for $z_{\rm acc, prim}$:
\begin{enumerate}
\item the highest accretion redshift, $z_{\rm acc,1}$, 
\item the most recent $z_{\rm acc, n}$, 
\item the highest accretion redshift that is not followed by a continuous period of more than two dynamical times as a host-halo before being re-accreted (or reaching z=0),
\item and the highest accretion redshift that occurs after $z_{\rm peak}$.  
\end{enumerate}

The first definition for $z_{\rm acc, prim}$ we examine is $z_{\rm acc,1}$.  We can eliminate this as a viable definition because we find that a significant fraction of host-haloes at $z\sim0$ were briefly identified as a sub-halo at high redshifts.  An example is shown in the left hand panel of Fig. \ref{fig:growth_hist}.  We find that between 4\% and 10\% of haloes more massive than $10^{12}h^{-1}M_{\odot}$ have $z_{\rm acc,1}$ and $z_{\rm ejt,1}$ $>3$ and remain a host-halo up to $z\sim0$.  It is clear that these haloes should be treated more as traditional host haloes than haloes that host satellite galaxies.   

The second definition we consider for $z_{\rm acc, prim}$ is the most recent accretion redshift.  This definition suffers the same problem as the previous by assigning too many host-haloes a high redshift $z_{\rm acc, prim}$.  In addition, such a definition ignores the accretion history of backsplash haloes by only considering the most recent accretion event.  Up to 60\% of sub-haloes are on orbits whose apocenter is beyond the virial radius of their effective host-halo, and around $\sim 10\%$ of accreted sub-structure is found beyond the virial radius of their associated host-halo at $z\sim 0$ \citep{Lin:2003bi, Gill:2005ge, Sales:2007kq, Ludlow:2009ix,  Wetzel:2014up, vandenBosch:2017fj}.  The middle panel of Fig. \ref{fig:growth_hist} shows an example of a halo which was accreted and ejected multiple times in line with the expectation for backslash haloes.  

To address both spurious high-redshift accretions and backsplash haloes at lower redshift, we consider a third definition for $z_{\rm acc, prim}$ that takes into account the amount of time a halo remains a host-halo after being ejected.  For backsplash haloes, the time-scale for re-accretion will be on the order of dynamical time.  If a halo remains a host-halo for much more time, evolution as a typical host-halo is more likely as in the case of host-haloes which were briefly identified as a sub-halo at high-redshift.

To this end, we calculate the number of dynamical times elapsed between redshift z and 0 as:
\begin{equation}
N_{\tau}(z) = \int_0^{t(z)} \frac{\mathrm{d}t}{\tau_{\rm dyn}(t)}
\label{eq:t_dyn}
\end{equation}
where $\tau_{\rm dyn}$ is the dynamical time given by:
\begin{align}
\tau_{\rm dyn}(t) &= \sqrt{\frac{3\pi}{16G\bar{\rho}_h(z)}} \\ \nonumber
                           &=1.628 ~h^{-1}{\rm Gyr}\left[ \frac{\Delta_{\rm vir}(z)}{178} \right] \left[ \frac{H(z)}{H_0}\right]^{-1}
\end{align}
where $\bar{\rho}_h(z)$ is the average density of a virialized dark matter halo at redshift $z$.  The number of dynamical times between the $i^{\rm th}$ ejection and the $i+1^{\rm th}$ accretion is then given by:
\begin{equation}
 \Delta N_{\tau} = N_{\tau}(z_{\rm ejt, i}) - N_{\tau}(z_{\rm acc, i+1})
\end{equation}
For any accretion redshift which is not followed by a continuous time, $\Delta N_{\tau}$, as a host-halo, we mark as the primary accretion redshift.  For haloes that remain a host for $\Delta N_{\tau}$ after being ejected, we disregard the previous accretion events when defining $z_{\rm acc, prim}$.  We find that $\Delta N_{\tau} = 2$ is a good threshold to separate backsplash-ing sub-haloes and host-haloes with spurious high redshift accretion events.  

The final definition for the primary accretion redshift we explore is the highest redshift accretion that occurs after $\widetilde{M}_{\rm peak}$.  This naturally removes any prior accretion events that were followed by mass growth while also generally picking out the initial accretion redshift for haloes that backsplash.  This definition lines up with $z_{\rm acc, 1}$ in the middle and right-hand panels of Fig. \ref{fig:growth_hist}.  We also find that $z_{\rm acc, prim}$ defined using the last two definitions (iii, iv) are different in less than 2\% of haloes with mass greater than $10^{12}h^{-1}M_{\odot}$.  Given the simplicity of this definition, we adopt this as our fiducial $z_{\rm acc, prim}$ in the rest of this paper and simply refer to it as $z_{\rm acc}$. 

\begin{figure}
\includegraphics[width=\columnwidth]{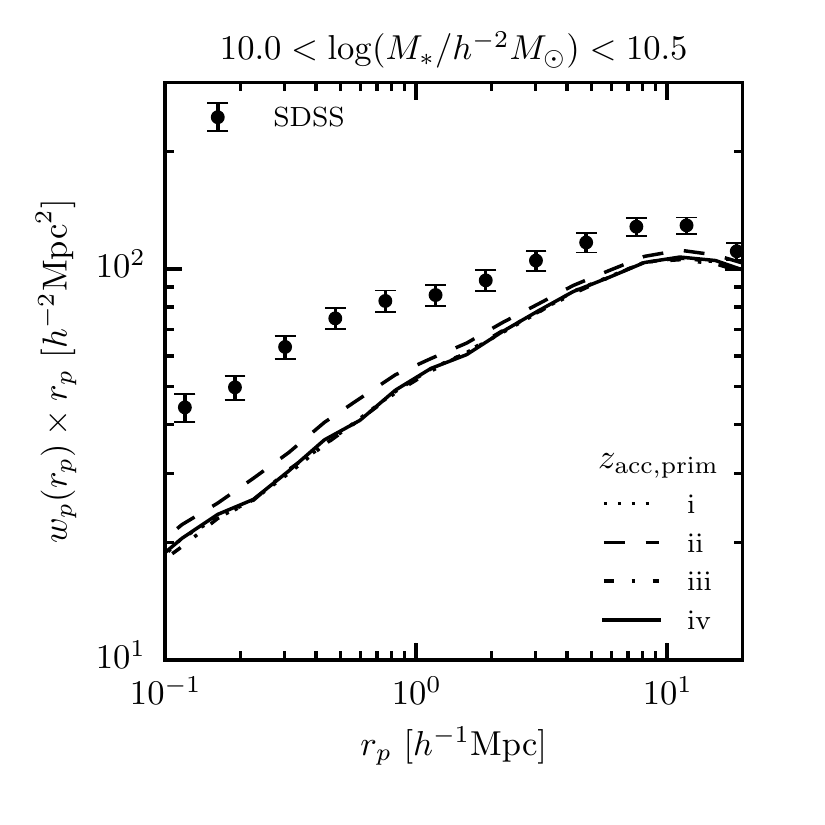}
\caption{Similar to Fig. \ref{fig:orphan_clustering_2}, a comparison of the effect of different definitions of $z_{\rm acc, prim}$ on $w_p(r_p)$ in the M13 model.  The model predictions for different $z_{\rm acc, prim}$ definitions are shown as lines of various styles corresponding to the definitions listed in Appendix \ref{sec:halo_acc_time}.  The lines for definitions i, iii, and iv fall nearly on-top of each other.}
\label{fig:zacc_wp_comp}
\end{figure}

We show the effect on clustering for different definitions of $z_{\rm acc, prim}$ for the M13 model in Fig. \ref{fig:zacc_wp_comp}.  The only significant difference is between the last accretion redshift (definition ii) and the others (i, iii, iv).  Using the last accretion redshift results in satellites with larger stellar masses relative to the other definitions as a result of the evolution in the SMHM relation towards larger stellar masses at fixed halo mass as $z \rightarrow 0$.  Satellites that are are ejected get a boost in stellar mass relative to those that remain satellites.  We consider this an unappealing model for satellite evolution.

\subsection{Halo Formation Time}

We calculate the formation time of a halo, $z_{f}$, as the redshift at which a halo is first identified as exceeding a mass larger than or equal to $f \times \widetilde{M}_{\rm peak}$ while not identified as a sub-halo.  In Fig. \ref{fig:growth_hist} we show $z_f$ for f=$0.1,0.2,...,0.9$ for three different halo growth histories marked with short vertical dashes below $m_{\rm vir}(z)$ in each panel.     

\section{Stellar Mass Conversions}
\label{appendix:stellar_mass_conv}

\begin{figure}
\includegraphics[width=\columnwidth]{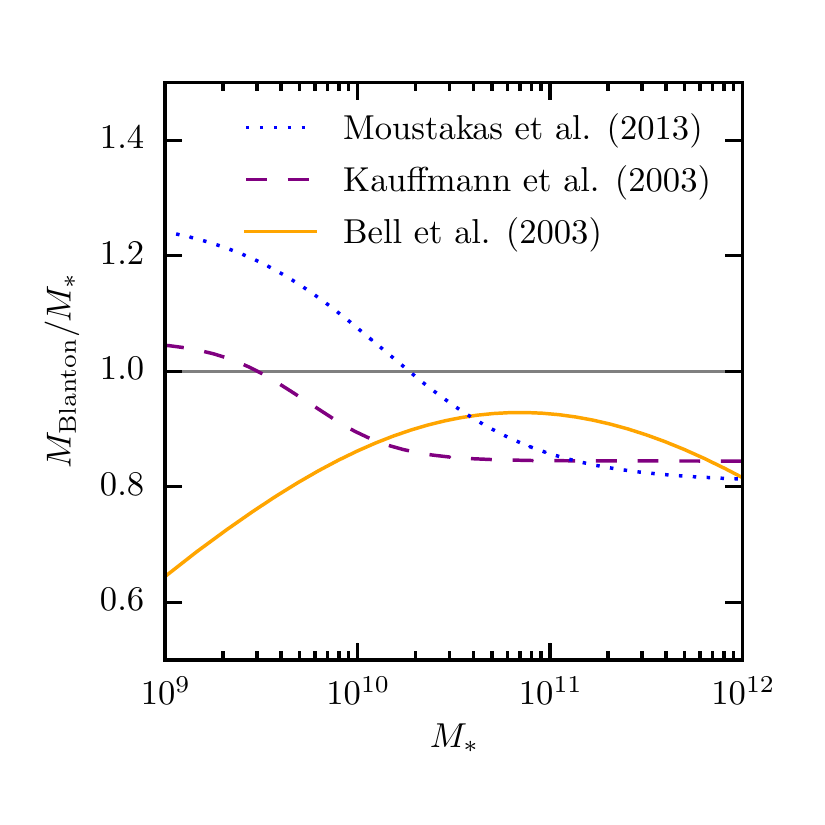}
\caption{mass conversion between Blanton stellar masses and the other stellar mass systems used in this paper.}
\label{fig:mstar_fconv}
\end{figure}

Each of the evolving models in this paper (M13, B13, and Y12) was tuned to fit stellar mass functions based on different methods to measure $M_*$.  In order to facilitate comparisons between models, we apply a set of simple conversion to correct for the mean difference in order to make the stellar masses more consistent with the \citet{Blanton:2007cv} stellar masses, $M_{\rm Blanton}$, used in LW09 and the RM and RV models.  A summary of these functions is shown in Fig. \ref{fig:mstar_fconv}.

Here we describe in detail each of these conversions.  M13 fits to the LW09 stellar mass function at $z \sim 0$ which was modified based on a conversion suggested by \citet{Guo:2010do} which transforms the stellar masses based on the SDSS r-band Petrosian magnitudes to ones based on SDSS r-band model magnitudes.  We undo this modification by reducing the stellar masses in M13 by $10\%$.  B13 fits to the \citet{Baldry:2008hm} and \citet{Moustakas:2013il} mass functions at $z < 0.2$.  The \citet{Baldry:2008hm} stellar masses are an average of many different techniques which makes a simple conversion prescription difficult.  On the other hand, \citet{Moustakas:2013il} provide a comparison between masses derived using the {\tt iSEDfit} and the \citet{Blanton:2007cv} masses.  We find the the mean difference is well fit by:
\begin{equation}
\log(M_{\rm Blanton}/M_{\rm iSEDfit}) = a_1+a_2\tanh\left(\frac{M_{\rm iSEDfit} - a_3}{a_4}\right)
\end{equation}
where $a_1 = 0.0056$, $a_2 = -0.098$, $a_3 =10.53$, and $a_4 = 0.82$.  We transform the B13 masses using this relation and find satisfactory results.  The Y12 model uses stellar masses based on the technique of \citet{Bell:2003hs} assuming a universal IMF \citep{Kroupa:2001ki, Borch:2006eu}.  We use the inverse of the transformation between \citet{Bell:2003hs} and \citet{Blanton:2007cv} masses provided in Appendix A in LW09:
\begin{align}
\log(M_{\rm Bell}/M_{\rm Blanton}) = a_1 &+ a_2 M_{\rm Blanton} + a_3 M_{\rm Blanton}^2 \nonumber \\
  & + a_3 M_{\rm Blanton}^3 + a_3 M_{\rm Blanton}^4
\end{align}
where $a_1 = 2.0$, $a_2 = -0.043$, $a_3 = -0.045$, $a_4 = 0.0032$, and $a_5=-2.1\times10^{-5}$.

\section{Orphan Galaxies}
\label{sec:clones}
 
We define ``orphan'' galaxies as satellites in our mock galaxy catalogues which have no identified sub-halo.  Our implementation of abundance matching requires a halo or sub-halo be associated with each galaxy.  In order to add the flexibility of including a population of orphan galaxies, we post-process the halo catalogues, adding ``clone'' sub-haloes that are made available to host orphan galaxies.

To generate a clone, we randomly draw from the list of all sub-haloes to choose a ``donor'' sub-halo.  The clone sub-halo receives all the properties of the donor sub-halo (e.g. $z_{\rm acc}$) except its phase space coordinates and those properties associated with it's host-halo.  A new host-halo is chosen for the clone by randomly choosing a host-halo with a mass close to that of the donor's host ($\pm 0.1~ {\rm dex}$).  We apply two methods to assign a new set of phase space coordinates to the clone:
\begin{enumerate}
\item maintaining the relative position, $\Delta \vec{x}$, and velocity, $\Delta \vec{v}$, between a clone's host-halo as in the donor's host-halo ({\tt sub-profile}),
\item assigning the phase-space coordinates of a randomly selected particle belonging to the clone's host to the clone ({\tt dm-profile}).
\end{enumerate}
For the former method, we use the positions and velocities of host-haloes and sub-haloes from the {\tt ROCKSTAR} halo catalogue.  For the later, we use a down-sampled catalogue of dark matter particles consisting of $\sim 1\%$ of all particles to facilitate computational ease.  We assign particles to host-haloes by finding all particles within a distance $r_{\rm vir}$ of each host.  When a particle can be assigned to more than one host under this condition, we assign the particle to the nearest host-halo.  After this process, we find that $\sim 1\%$ of clones occupy a host-halo with no associated particles (in our down-sampled catalogue).  In this small fraction of cases, we revert to the first method.   

\begin{figure}
\includegraphics[width=\columnwidth]{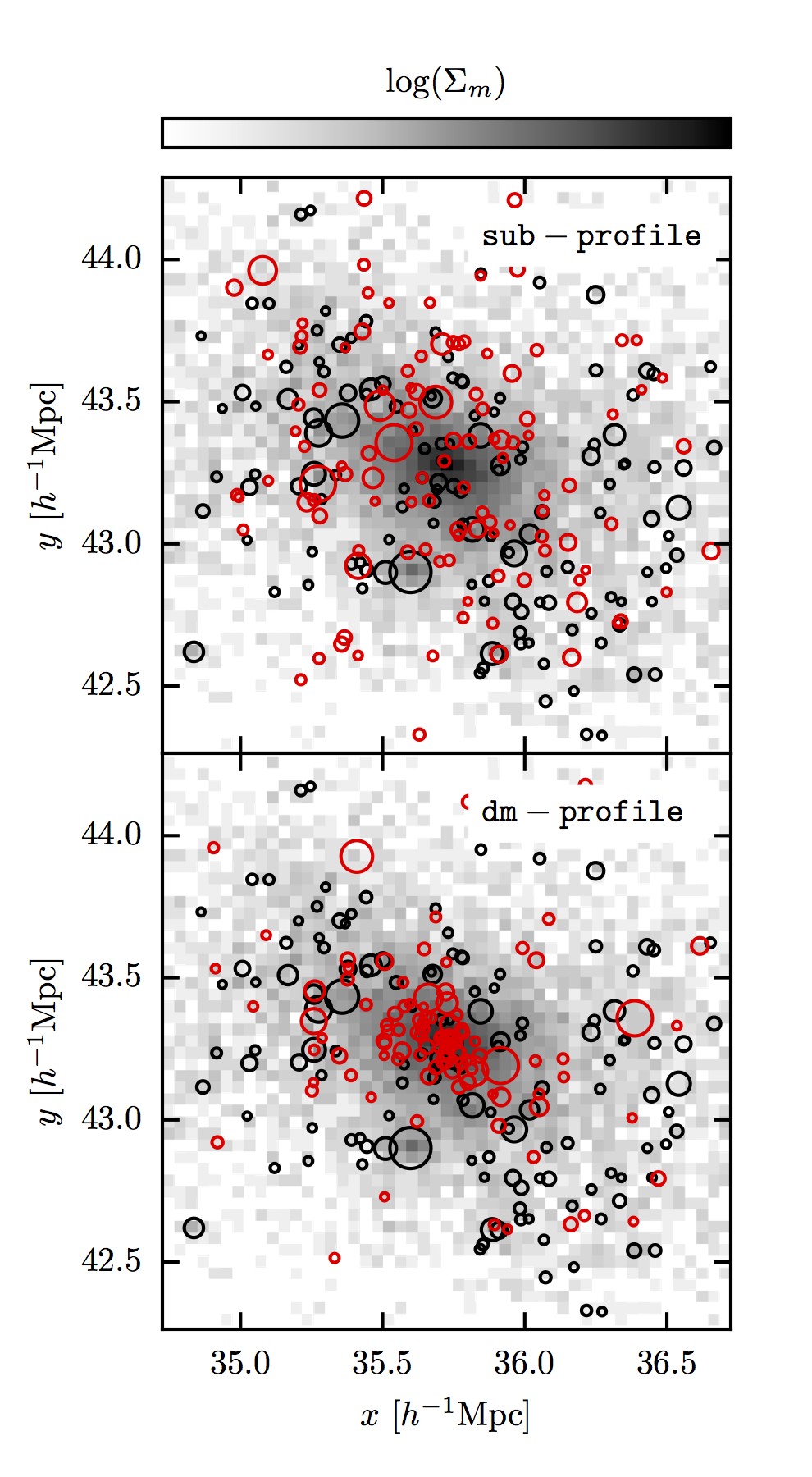}
\caption{The projected distribution of sub-haloes with $M_{\rm peak}>10^{11}~h^{-1}M_{\odot}$ (black circles) for an example host-halo ($m_{\rm vir}\simeq 10^{14}~h^{-1}M_{\odot}$), where the projected density of dark matter is shown in grey-scale (the same in upper and lower panels).  In addition, the upper panel shows clone sub-haloes (red circles) where the relative distance and velocity with respect to the centre of mass is preserved ({\tt sub-profile}).  The bottom panel shows clone sub-haloes where the position and velocity is assigned by drawing random particles from the host-halo ({\tt dm-profile}).  The size of the circles is proportional to the viral radius at peak mass, $r_{\rm vir}(z_{\rm peak})$, of each sub-halo.}
\label{fig:subhalo_positions}
\end{figure}

Each of these methods has merits.  The first method acts under the assumption that the sub-haloes that host orphan galaxies (and missing from the available halo catalogues) are a fair sampling of all sub-haloes.  These clones will have the same radial profile as typical sub-haloes within host-haloes of equivalent mass.  The second method results in a more centrally concentrated population of orphans, one that also naturally follows the shape of the host-halo.  This may be a more appropriate if the majority of orphans occupy sub-haloes that are missing because they are hard to identify in the dense central regions of host-haloes or highly evolved sub-haloes which have sunk to the central regions of their host-halo.  Neither of these methods will preserve sub-halo-sub-halo correlations.  In particular, neither of these methods specifically deals with higher-order sub-haloes, e.g. sub-sub-haloes, and treats all sub-haloes regardless of order the same in the cloning process.  To help with visualization of each of these methods, we plot the position of sub-haloes and clones in Fig. \ref{fig:subhalo_positions} for both methods.  In the bottom panel, one can see that the clones are more centrally concentrated than both the extant sub-haloes and clones assigned positions maintaining the relative distance to the host-halo centre. 

\section{Degrading Rank Order Correlations}
\label{appendix:ranks}

Given data sets that contain finite realizations, e.g. $x_1, x_2, ..., x_n$, and $y_1, y_2, ..., y_n$ of length $n$, of two random variables $X$ and $Y$, the correlation between the variables can be characterized by the Spearman's rank order correlation coefficient:
\begin{equation}
\rho_{XY} =1-\frac{6\, \sum_i^{n} d_i^2}{n(n^2-1)}
\end{equation}
where $d_i$ is the difference in ranks of $x_i$ and $y_i$:
\begin{equation}
d_i = n_i^x- n_i^y
\end{equation}
For example, if $X$ and $Y$ are both in rank order, e.g. $n_1^x = 1, n_2^x=2, ..., n_n^x = n$, then $\rho_{XY}=1.0$.

Here we describe an algorithm to degrade the ordering of two variables, $X$ and $Y$, each of length $n$.  To begin with a positive correlation, each variable is placed in rank order:  
\begin{align}
X^{\prime} &= {\rm RANK}(X,X) \nonumber \\
Y^{\prime} &= {\rm RANK}(Y,Y) 
\end{align}
where the ${\rm RANK(A,B)}$ operator sorts $A$ by the rank order values of $B$.  It should be noted that to begin with a negative correlation between $X$ and $Y$, $X^{\prime}$ would be put in inverse rank order, i.e. $X^{\prime} = {\rm RANK}(X,-X)$.  Next, a new variable, $Q$, is calculated for $X^{\prime}$ from the ranks by adding a normal random variable to each rank:   
\begin{equation}
q_i = n^{x^{\prime}}_i + \mathcal{N}(0, \sigma_q \times n)
\end{equation}
where $\sigma_q$ is approximately the standard deviation of the change in the order relative to the length of $n$.  $X^{\prime}$ can then be re-ordered by $Q$:
\begin{equation}
X^{\prime\prime}  = {\rm RANK}(X^{\prime}, Q)
\end{equation}
Henceforth, we will refer to these two variables with transformed ordering simply as $X$ and $Y$.

\begin{figure}
\includegraphics[width=\columnwidth]{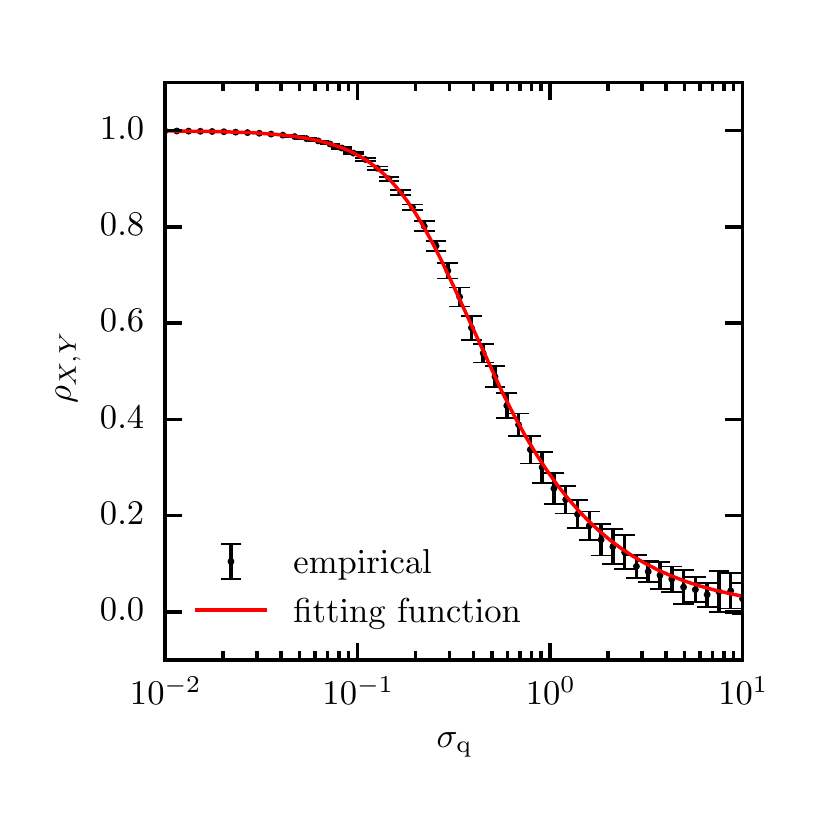}
\caption{The relation between rank scatter parameter, $\sigma_q$, and the Spearman's correlation coefficient, $\rho$.  The fitting function, eq. \ref{eq:rho_fit_func}, is shown as a red line.}
\label{fig:rho_sigma}
\end{figure}

This method is inherently random in nature, and given a value of $\sigma_q$, the rank order correlation between $X$ and $Y$ will vary depending on the size of the data sets.  In Fig. \ref{fig:rho_sigma} we examined the relation between the correlation coefficient, $\rho_{XY}$, and $\sigma_q$ for two uniform random variables of length $n=10^{3}$.  For each value of $\sigma_q$ we repeat the process described above 100 times.  The error bars in Fig. \ref{fig:rho_sigma} are the standard deviation in $\rho_{XY}$ from these 100 realizations.  From this, we derive the relation between $\langle \rho_{XY} \rangle$ and $\sigma_q$, and we use this relation to choose $\sigma_q$ for a desired value of $\rho_{XY}$.  We provide an accurate fitting function for the relation given by:
\begin{align}
\label{eq:rho_fit_func}
&\langle \rho_{XY} \rangle(\sigma_q) = 1-f(\sigma_q) \\
&f(x) = \frac{1}{2}e^{-\left(\frac{x}{x_1}\right)^{\alpha}} + \frac{1}{2} \left[1+ \left( \frac{x}{x_2} \right)^{\beta} \right]^{-1} \nonumber
\end{align}
where $x_1=0.650$, $x_2=0.302$, $\alpha=-1.067$, and $\beta=-1.978$.  Furthermore, this relation is independent of the size of $X$ and $Y$ and is not affected by the distribution of values themselves since it is based on the rank ordering.

%

% Don't change these lines
\bsp	% typesetting comment
\label{lastpage}
\end{document}